\documentclass{jpp-AAS}

\usepackage{graphicx}
\usepackage{natbib}

\usepackage{amsmath}
\usepackage{amssymb}
\usepackage[makeroom]{cancel}
\usepackage{color}
\usepackage[colorlinks=true,linkcolor=blue,citecolor=blue]{hyperref}

\ifCUPmtlplainloaded \else
  \checkfont{eurm10}
  \iffontfound
    \IfFileExists{upmath.sty}
      {\typeout{^^JFound AMS Euler Roman fonts on the system,
                   using the 'upmath' package.^^J}%
       \usepackage{upmath}}
      {\typeout{^^JFound AMS Euler Roman fonts on the system, but you
                   dont seem to have the}%
       \typeout{'upmath' package installed. JPP.cls can take advantage
                 of these fonts, if you use 'upmath' package.^^J}%
      }
  \else
  \fi
\fi

\ifCUPmtlplainloaded \else
  \checkfont{msam10}
  \iffontfound
    \IfFileExists{amssymb.sty}
      {\typeout{^^JFound AMS Symbol fonts on the system, using the
                'amssymb' package.^^J}%
       \usepackage{amssymb}%
       \let\le=\leqslant  
       \let\ge=\geqslant  
      }{}
  \fi
\fi

\ifCUPmtlplainloaded \else
  \IfFileExists{amsbsy.sty}
    {\typeout{^^JFound the 'amsbsy' package on the system, using it.^^J}%
     \usepackage{amsbsy}}
    {\providecommand\boldsymbol[1]{\mbox{\boldmath $##1$}}}
\fi

\newcommand{\figref}[1]{figure~\ref{#1}}
\newcommand{\figsref}[1]{figures~\ref{#1}}
\newcommand{\Figref}[1]{Figure~\ref{#1}}

\newcommand{\secref}[1]{section~\ref{#1}}

\newcommand{\secsand}[2]{sections~\ref{#1} and \ref{#2}}

\newcommand{\apref}[1]{appendix~\ref{#1}}

\renewcommand{\eqref}[1]{equation~(\ref{#1})}

\newcommand{\exref}[1]{(\ref{#1})}
\newcommand{\exsdash}[2]{(\ref{#1}--\ref{#2})}

\newcommand{\bea}{\begin{eqnarray}}
\newcommand{\eea}{\end{eqnarray}}
\newcommand{\beq}{\begin{equation}}
\newcommand{\eeq}{\end{equation}}
\newcommand{\lt}{\left}
\newcommand{\rt}{\right}

\newcommand{\la}{\langle}
\newcommand{\ra}{\rangle}

\newcommand{\mbf}[1]{\boldsymbol{#1}}

\newcommand{\dd}{\partial}
\newcommand{\vdel}{\mbf{\nabla}}

\newcommand{\const}{\mathrm{const}}

\newcommand{\vr}{\mbf{r}}

\newcommand{\kperp}{k_\perp}

\newcommand{\kpar}{k_\parallel}

\newcommand{\tnl}{t_\mathrm{nl}}
\newcommand{\tc}{t_\mathrm{c}}
\newcommand{\tcoll}{t_\nu}

\newcommand{\vv}{\mbf{v}}

\newcommand{\vperp}{v_\perp}
\newcommand{\vpar}{v_\parallel}

\newcommand{\vth}{v_{{\rm th}}}
\newcommand{\vths}{v_{{\rm th}s}}
\newcommand{\vthi}{v_{{\rm th}i}}

\newcommand{\vthe}{v_{{\rm th}e}}

\newcommand{\ope}{\omega_{\mathrm{p}e}}
\newcommand{\lDe}{\lambda_{\mathrm{D}e}}

\newcommand{\EE}{{\mathfrak E}}

\newcommand{\vE}{\mbf{E}}
\newcommand{\vB}{\mbf{B}}

\newcommand{\dn}{\delta n}

\newcommand{\dnb}{\delta\bar{n}_e}

\newcommand{\GG}{G}
\newcommand{\tf}{\tilde f}

\renewcommand{\tt}{\tilde t}

\newcommand{\kap}{\varkappa}
\newcommand{\al}{\lambda}
\newcommand{\be}{\mu}
\newcommand{\kk}{\kappa}
\newcommand{\xx}{\xi}
\newcommand{\yy}{\eta}

\newcommand{\ephi}{\varphi}
\newcommand{\bphi}{\bar{\phi}}

\newcommand{\uu}{u}

\newcommand{\mmin}{m_0}
\newcommand{\smin}{s_0}
\newcommand{\tmin}{\tau_0}
\newcommand{\tmax}{\tau_\mathrm{max}}

\newcommand{\FM}{F_{\mathrm{M}}}
\newcommand{\df}{\delta\!f}

\newcommand{\rmd}{\mathrm{d}}

\renewcommand{\Re}{\mathrm{Re}}
\renewcommand{\Im}{\mathrm{Im}}

\newcommand{\underbox}[3]{\underbrace{#1}_{\parbox{#2}{\begin{center}\small #3\end{center}}}}

\title[Vlasov turbulence in Fourier--Hermite space]{A solvable model of Vlasov-kinetic plasma 
turbulence in Fourier-Hermite phase space}
\author[T.\ Adkins and A.\ A.\ Schekochihin]%
{T.~Adkins%
\thanks{Email: toby.adkins@merton.ox.ac.uk} 
and A.~A.~Schekochihin%
\thanks{Email: alex.schekochihin@physics.ox.ac.uk}}  
\affiliation{
Rudolf Peierls Centre for Theoretical Physics, University of Oxford, 1 Keble Road,\\ 
Oxford OX1 3NP, UK
\\[\affilskip]
Merton College, Merton Street, Oxford OX1 4JD, UK}

\begin{document}

\maketitle

\begin{abstract}
A class of simple kinetic systems is considered, described by the 1D Vlasov--Landau equation 
with Poisson or Boltzmann electrostatic response and an energy source. 
Assuming a stochastic electric field, a solvable model is constructed for 
the phase-space turbulence of the 
particle distribution. The model is a kinetic analog of the Kraichnan--Batchelor 
model of chaotic advection. The solution of the model is found in Fourier--Hermite 
space and shows that the free-energy flux from low to high Hermite moments is 
suppressed, with phase mixing cancelled on average by anti-phase-mixing (stochastic 
plasma echo). This implies that Landau damping is an ineffective route to dissipation 
(i.e., to thermalisation of electric energy via velocity space).
 The full Fourier--Hermite spectrum is derived. Its asymptotics 
are $m^{-3/2}$ at low wave numbers and high Hermite moments ($m$) and 
$m^{-1/2}k^{-2}$ at low Hermite moments and high wave numbers ($k$). 
These conclusions hold 
at wave numbers below a certain cut off (analog of Kolmogorov scale), which increases 
with the amplitude of the stochastic electric field and scales as inverse square 
of the collision rate. The energy distribution 
and flows in phase space are a simple and, therefore, useful example of competition between 
phase mixing and nonlinear dynamics in kinetic turbulence, reminiscent of more realistic but  
more complicated multi-dimensional systems that have not so far been amenable to 
complete analytical solution.   
\end{abstract}

\section{Introduction} 

One of the most distinctive properties of a weakly collisional plasma as a 
physical system is the intricate phase-space dynamics associated with the 
interaction between electromagnetic fields and charged particles. 
The signature plasma-physics phenomenon of \citet{landau46} damping 
consists essentially in the removal of free energy from an electromagnetic 
perturbation and its transfer ``into phase space'', i.e., into fine-scale 
structure of the perturbed distribution function in velocity space 
(``phase mixing''). It has long been realised that nonlinear effects can lead 
to Landau damping shutting down, both for broad-spectrum fields and individual 
monochromatic waves \citep{vedenov62,oneil65,mazitov65,manheimer68,weiland92}, 
or even to apparently damped perturbations coming back from phase space, 
a phenomenon called ``plasma echo'' \citep{gould67,malmberg68}. 
Fast-forwarding over several decades of plasma-turbulence 
theory (see \citealt{krommes15} and \citealt{laval16} for review and references), 
the notion of ``phase-space turbulence'', pioneered by \citet{dupree72}, 
has, in the recent years, again 
become a popular object of study, treated either, Dupree-style, in 
terms of formation of phase-spaces structures and their 
effect on the transport properties of the plasma \citep{kosuga11,kosuga14,kosuga17,lesur14a,lesur14b}
or in terms of a kinetic cascade carrying free energy to collisional scales 
in velocity space \citep{watanabe04,sch08,sch09,tatsuno09,plunk10,plunk11,banon11prl,teaca12,teaca16,hatch14,kanekar15phd,sch16,parker16,servidio17}. 

Within the latter strand, 
a direct precursor to the present study is the paper by \citet{sch16}, 
who proposed, using electrostatic 
drift-kinetic turbulence as a prototypical example of kinetic turbulence, 
that a key effect of nonlinearity on phase-space dynamics would be 
an effective suppression of Landau damping---meaning that  
the free-energy flux from small to large scales in velocity space 
associated with stochastic echos (``anti-phase-mixing'', or ``phase unmixing'') 
largely cancels the phase-mixing flux (``Landau damping'') from large to small scales. 
A signature of this effect is a Hermite spectrum of free energy that 
is steeper for the nonlinear, turbulent perturbations than for the linear, 
Landau-damped ones (seen in numerical simulations of \citealt{watanabe04}, 
\citealt{hatch14} and \citealt{parker16}). As a result,
the limit of vanishing collisionality ceases to be a singular limit 
at long wavelengths (as it is in the linear regime; see, e.g., \citealt{kanekar15}) 
and most of the entropy production occurs below the Larmor scale 
(i.e., outside the drift-kinetic approximation). 

While some theoretical predictions of \citet{sch16} appear 
to have found a degree of numerical backing \citep{parker16},
their theory of stochastic echo was a qualitative one, relying 
on plausible scaling arguments, rather like the theory of hydrodynamic 
turbulence mostly does to this day \citep{davidson04}. 
In addition to such arguments, understanding of fluid turbulence 
has benefited greatly from the development of simplified 
solvable models, the most famous of which is the ``passive-scalar'' 
model describing the behaviour of a scalar field chaotically advected 
by an externally determined random flow \citep{kraichnan68,kraichnan74,kraichnan94,falkovich01}. 
Under certain assumptions about the nature of this flow, 
it is possible to solve for the passive-scalar statistics analytically, 
leading to a number of interesting and nontrivial predictions, some of which 
appear to carry over qualitatively or even quantitatively to more realistic turbulent 
systems and all of which have proved stimulating to turbulence 
theorists. In view of this experience, seeking a maximally simplified but 
analytically solvable model appears to be worthwhile.   

In this paper, we propose such a solvable model, based on the 
much-studied 1D Vlasov--Poisson system. The phase space for this 
system is two-dimensional (one spatial and one velocity coordinate). 
The particle distribution function in a turbulent state can be 
described in terms of its Fourier--Hermite spectrum.  
We show that, given a stochastic electric field, 
the only physically sensible solution features zero net free-energy flux 
from low (``fluid'') to high (``kinetic'') Hermite moments, meaning that 
the Landau damping is suppressed and the low moments are 
energetically insulated from the rest of the phase space. 
The underlying mechanism of this suppression is the stochastic-echo effect. 
The resulting Hermite spectrum is, asymptotically, 
$m^{-3/2}$ at large Hermite orders $m$ (compared to $m^{-1/2}$ 
for linear Landau damping; see \citealt{zocco11} and \citealt{kanekar15}) 
and so the limit of small collisionality is nonsingular (collisional 
dissipation tends to zero if the collision rate does). The 
corresponding Fourier spectrum of the low-$m$ Hermite moments is $m^{-1/2}k^{-2}$. 
Surveyed over the entire phase space, the 
phase-mixing (Landau-damped) and anti-phase-mixing (echoing) 
components of the distribution function have an interesting and 
not entirely trivial self-similar structure, which can nevertheless be fully extracted 
analytically and bears some resemblance to what is seen 
in various numerical simulations. 
A finite collision rate imposes a finite wave-number cutoff on this solution, 
which is the analog of the Kolmogorov scale for the Vlasov-kinetic turbulence.  

The rest of the paper is organised as follows. In \secref{sec:model}, we describe 
a family of plasma systems that can be reasonably modelled by the equations 
studied in this paper (electron Langmuir turbulence, ion-acoustic turbulence, 
Zakharov turbulence). In \secref{sec:formalism}, we recast these equations in 
Fourier--Hermite space and introduce the formalism within which the phase mixing 
anti-phase-mixing can be studied explicitly (this formalism is similar to 
one developed by \citealt{sch16}, but with minor adjustments). 
In \secref{sec:method}, we make the 
approximations required to render the problem solvable and derive an equation for 
the Fourier--Hermite spectrum of the distribution function. In \secref{sec:solution}, 
we solve this equation, obtaining the results promised above (a qualitative summary 
of this solution and an assessment of the effect of finite collisionality on it 
are given in \secref{sec:colls}; a quicker, but perhaps 
less mathematically complete route to it than one pursued in the 
main text is described in \apref{ap:ssim}). Finally, results are summarised 
and limitations, implications, applications and future directions discussed 
in \secref{sec:disc}. 

\section{Models: Vlasov--Poisson system and its cousins}
\label{sec:model}

The standard Vlasov--Poisson system describes a plasma in the absence 
of magnetic field. For each species ($s=e$ electrons or $s=i$ ions), 
the distribution function obeys the Vlasov--Landau equation 
\beq
\frac{\dd f_s}{\dd t} + \vv\cdot\vdel f_s 
- \frac{q_s}{m_s}(\vdel\phi)\cdot\frac{\dd f_s}{\dd\vv} = \lt(\frac{\dd f_s}{\dd t}\rt)_\mathrm{c},
\label{eq:Vlasov}
\eeq
where $q_s$ and $m_s$ are the charge and mass of the particles, 
the term in the right-hand side is the collision operator, and $\phi$ is the electrostatic 
potential, satisfying Poisson's equation 
\beq
-\nabla^2\phi = 4\pi\sum_s q_s\int\rmd^3\vv\,\df_s.
\label{eq:Poisson}
\eeq 
We are formally splitting the distribution function into mean and perturbed parts,
\beq
f_s = f_{0s}(\vv) + \df_s(t,\vr,\vv),
\eeq
where $f_{0s}$ is spatially homogeneous and we assume that there is no 
mean electric field. Only $\df_s$ enters the Poisson equation \exref{eq:Poisson}
because the plasma is neutral on average. We do not 
require that $\df_s\ll f_{0s}$ everywhere, although we do assume that 
any temporal evolution of the mean distribution is slow compared 
to that of the perturbation. We take the mean distribution to be 
Maxwellian, 
\beq
f_{0s} = \frac{n_{0s}}{(\pi\vths^2)^{3/2}} e^{-|\vv|^2/\vths^2},
\label{eq:Maxwellian}
\eeq 
where $\vths= \sqrt{2T_s/m_s}$ is the thermal speed of the particles of species $s$ 
and $n_{0s}$ and $T_s$ are their number density and temperature, respectively.  

Several simplified models can be constructed, leading to mathematically 
similar sets of equations. 

\subsection{Electron Vlasov--Poisson plasma}
\label{sec:emodel}

Assuming cold ions, or, equivalently, restricting our consideration to perturbations 
with frequencies of the order of the electron plasma frequency, 
\beq
\omega\sim\ope = \sqrt{\frac{4\pi e^2 n_{0e}}{m_e}}, 
\eeq
where $-e\equiv q_e$, we may set
$\df_i=0$. Further restricting our consideration to a single spatial dimension, 
$x$, we have $\df_e = \df_e(x,\vv)$. We may now introduce 
the following reduced, non-dimensionalised fields and variables:
\beq
g(x,v) = \frac{\vthe}{n_{0e}}\int\rmd v_y\rmd v_z \df_e,\quad
\FM(v) = \frac{1}{\sqrt{\pi}}\,e^{-v^2},\quad
v = \frac{v_x}{\vthe},\quad
\ephi = -\frac{e\phi}{T_e}.  
\label{eq:edefs}
\eeq 
We may also non-dimensionalise $t\ope \to t$ and $x/\sqrt{2}\,\lDe \to x$,
where $\lDe = \vthe/\sqrt{2}\,\ope$ is the electron Debye length. 
In this notation, the Vlasov--Poisson system becomes 
\beq
\frac{\dd g}{\dd t} + v\frac{\dd g}{\dd x} + v \FM \frac{\dd\ephi}{\dd x} 
- \frac{1}{2}\frac{\dd\ephi}{\dd x}\frac{\dd g}{\dd v} = C[g],
\label{eq:g}
\eeq
\beq
\ephi = \alpha \int\rmd v\, g + \chi,
\label{eq:phi}
\eeq
where $-\alpha$ is twice the inverse Laplacian operator, $\alpha_k = 2/k^2$ in Fourier space.
We have added an ``external'' potential $\chi$ 
to represent energy injection in an analytically convenient fashion
(it will also acquire concrete physical meaning in \secsand{sec:Zmodel}{sec:stoch_acc}). 
Finally, we make a further simplification by using the \citet{lenard58} collision operator 
\beq
C[g] = \nu \frac{\dd}{\dd v}\lt(\frac{1}{2}\frac{\dd}{\dd v} + v\rt) g,
\label{eq:LB}
\eeq
with the proviso that it must be adjusted to conserve momentum and energy. 
This will not be a problem as the collision frequency $\nu$ will always be assumed 
small and so will only matter for the part of $g$ that varies quickly with $v$.

Besides being interesting in itself, the 1D Vlasov--Poisson system \exsdash{eq:g}{eq:phi} 
is an appealing minimal model that contains all the ingredients necessary for 
phase-space turbulence featuring a competition between phase mixing 
and nonlinearity.  

\subsection{Ion-acoustic turbulence}
\label{sec:imodel}

Another, mathematically similar, model describes  
electrostatic perturbations at low frequencies, where 
it is ion kinetics that matter, namely, 
\beq
\omega\sim k\vthi.
\eeq  
Since $m_e\ll m_i$ and assuming $T_i\sim T_e$, the electrons' velocities 
are $(m_i/m_e)^{1/2}$ larger than the ions' and so the 
kinetic equation \exref{eq:Vlasov} for $s=e$ becomes, on ion time scales,  
\beq
\vv\cdot\vdel f_e 
+ \frac{e}{m_e}(\vdel\phi)\cdot\frac{\dd f_e}{\dd\vv} = \lt(\frac{\dd f_e}{\dd t}\rt)_\mathrm{c}.
\eeq
This is solved by the Maxwell--Boltzmann distribution 
\beq
f_e = \frac{n_{0e}}{(2\pi T_e/m_e)^{3/2}}\exp\lt[-\frac{1}{T_e}\lt(\frac{m_e|\vv|^2}{2}-e\phi\rt)\rt].
\eeq
The electrons, therefore, have a Boltzmann response:  
\beq
n_e = n_{0e} e^{e\phi/T_e} \approx n_{0e}\lt(1 + \frac{e\phi}{T_e}\rt),
\label{eq:ne}
\eeq 
assuming $e\phi/T_e\ll1$. 
If $k\lDe\ll1$, the Poisson equation \exref{eq:Poisson} turns 
into the quasineutrality constraint: 
\beq
\frac{\dn_i}{n_{0i}} = \frac{\dn_e}{n_{0e}} = \frac{e\phi}{T_e}, 
\label{eq:quasineutrality}
\eeq
where the last equality follows from \exref{eq:ne} and the ion density 
perturbation has to be calculated from the perturbed ion distribution 
function, $\dn_i = \int\rmd^3\vv\, \df_i$. Restricting consideration again to 
1D perturbations, $\df_i = \df_i(x,\vv)$, and defining 
\beq
g(x,v) = \frac{\vthi}{n_{0i}}\int\rmd v_y\rmd v_z \df_i,\quad
\FM(v) = \frac{1}{\sqrt{\pi}}\,e^{-v^2},\quad
v = \frac{v_x}{\vthi},\quad
\ephi = \frac{Ze\phi}{T_i}, 
\label{eq:idefs}
\eeq
where $Ze \equiv q_i$, we find that $g$ again satisfies \exref{eq:g}. 
Using \exref{eq:quasineutrality} and again adding an external forcing $\chi$, we have
\beq
\ephi - \chi = \frac{ZT_e}{T_i} \frac{\dn_i}{n_{0i}} = \frac{ZT_e}{T_i} \int \rmd v\,g
\equiv \alpha \int \rmd v\,g. 
\label{eq:iphi}
\eeq
This is the same as \exref{eq:phi}, except now $\alpha = ZT_e/T_i$ is 
a constant rather than a differential operator. 

Note that the spatial and temporal coordinates in \exref{eq:g} can now be normalised 
$x/L \to x$ and $t\vthi/L\to t$ with an entirely arbitrary scale $L$ because 
the fundamental dynamics described by the ion equations---(damped) sound waves---do not 
have a special length scale. 

\subsection{Zakharov turbulence}
\label{sec:Zmodel}

Models allowing perturbations only on electron or only on ion 
scales are, in fact, of limited relevance to real plasma turbulence because 
interactions of Langmuir waves will promote coupling to low-frequency 
modes at the ion scales while those low-frequency modes will locally 
alter the plasma frequency, giving rise to a ``modulational'' nonlinearity 
in the electron-scale dynamics. Such a ``two-scale'' system has been extensively 
studied, mostly using the so-called \citet{zakharov72} equations, 
or, to be precise, a version of them in which both the electron and ion densities 
obey fluid-like equations (see reviews by
\citealt{thornhill78}, \citealt{rudakov78}, \citealt{goldman84}, 
\citealt{zakharov85}, \citealt{musher95}, \citealt{tsytovich95}, 
\citealt{robinson97} and \citealt{kingsep04}, 
of which the first and the last are the most readable) 
For electrons, the fluid approximation requires $k\lDe\ll1$ 
and for ions, $T_i\ll T_e$, so neither electron nor ion Landau damping 
is then important (because the phase velocities of the Langmuir and sound 
waves are much greater than $\vthe$ and $\vthi$, respectively). 

In a traditional 
approach to plasma turbulence, turbulence is what occurs at 
the scales (and in parameter regimes) where 
the dominant interactions are between wave modes (e.g., Langmuir or sound), 
which conserve fluctuation energy and transfer it, via a ``cascade'', to scales 
where waves can interact with particles---usually via Landau damping, 
linear and/or nonlinear. The latter processes are expected to lead to absorption 
of the wave energy by particles, i.e., its conversion into heat. Thus, 
regimes and scale ranges in which kinetic physics matters are viewed 
as dissipative (analogous to viscous scales in hydrodynamic turbulence). 
 
If one is not committed to such a dismissive attitude to kinetics 
in the way that a turbulence theorist in search of a fluid model might be, 
one may wish to explore how Zakharov's turbulence interfaces with 
the phase space. While the fluid approximation for electrons at 
long wave lengths ($k\lDe\ll1$) is sensible, the assumption of 
cold ions and hence unfettered sound propagation is fairly restrictive, 
so one may wish to remove it. It is then possible to derive a kinetic 
version of Zakharov's equations (as \citealt{zakharov72} in fact did), 
in which the ion kinetic equation 
stays intact [this is \exref{eq:Vlasov} with $s=i$ and $q_s=Ze$], but 
the potential $\phi$ in this equation is the ion-time-scale ($\sim1/k\vthi$) 
averaged potential---a kind of mean field against the background 
of electron-time-scale Langmuir oscillations. This mean potential is determined from 
the Poisson equation, which, since $k\lDe\ll1$, again takes the form 
of the quasineutrality constraint \exref{eq:quasineutrality}, but the 
ion-time-scale electron-density perturbation now contains both the 
Boltzmann response and the so-called ponderomotive one---essentially 
an effective pressure due to the average energy density of the 
Langmuir oscillations: 
\beq
\frac{\dn_i}{n_{0i}} = \frac{\dnb}{n_{0e}} 
= \frac{e\bphi}{T_e} - \frac{\overline{|\EE|^2}}{8\pi n_{0e} T_e}, 
\label{eq:Z_pmforce}
\eeq
where overbars denote averages over the electron time scales and 
$\EE$ is the electric field associated with the Langmuir waves. 
The resulting ion equations are the same as those derived 
in \secref{sec:imodel}, viz., the kinetic equation \exref{eq:g} with the definitions 
\exref{eq:idefs} (but $\phi\to\bphi$) 
and $\ephi$ given by \exref{eq:iphi}, or, equivalently, by \exref{eq:phi}, 
but with the ``external'' forcing now having a concrete physical meaning:
\beq
\chi = \frac{\overline{|\EE|^2}}{8\pi n_{0i} T_i}. 
\eeq
This forcing is, in fact, not independent of either $\ephi$ or $\df_i$, 
as $\EE$ satisfies a ``fluid'' equation for the Langmuir oscillations 
with the plasma frequency modulated by $\dnb$ (see \citealt{zakharov72} or 
any of the reviews cited above; a systematic derivation 
of Zakharov's equations from kinetics, which is surprisingly difficult 
to locate in the literature, can be found in \citealt{KTnotes}). 

Thus, yet again, we have a system of equations that is mathematically 
similar to the 1D Vlasov--Poisson system \exref{eq:g} and \exref{eq:phi}. 

\subsection{Stochastic-acceleration problem}
\label{sec:stoch_acc}

A simpler problem than the three preceding ones is to consider a population 
of ``test particles'', embedded in an externally imposed, statistically stationary  
stochastic electric field $\vE = -\vdel\phi$, and seek these particles' 
distribution function. It satisfies Vlasov's equation \exref{eq:Vlasov}, 
with collision integral now omitted. It can be restricted to 1D either by 
fiat or by considering particles in a magnetic field being accelerated by 
fast electric fluctuations parallel to it. 
In the limit of the field $E$ having a short correlation time compared 
to the characteristic time for particles to become trapped in the potential 
wells associated with $E$, the particles' spatially averaged distribution function 
is easily shown to satisfy a diffusion equation \citep{sturrock66}:\footnote{This 
is done entirely analogously to the calculation in \secref{sec:KK}, where 
the white-noise model for $\ephi$ is introduced and used \citep[cf.][]{cook78}.}
\beq
\frac{\dd f_0}{\dd t} = D\,\frac{\dd^2 f_0}{\dd v^2},\quad
D = \frac{e^2}{m^2}\int_0^\infty\rmd \tau \la E(t) E(t-\tau)\ra,
\label{eq:stoch_heating}
\eeq
where the electric-field correlation function should generally speaking be 
taken along particles' trajectories, but, in the limit of short correlation times, 
it is the same as the Eulerian correlation function. 

The solution of 
\exref{eq:stoch_heating} (assuming an initial $\delta$-shaped particle distribution) 
is a 1D Maxwellian with $\vth^2 = 4Dt$, expressing gradual secular heating 
of the test-particle population. At long times, this evolution can be treated as slow compared 
to the evolution of the perturbation $\df$ and so the latter considered to 
evolve against the background of a quasi-constant Maxwellian equilibrium. 
With the same normalisations as in \secref{sec:emodel}, the perturbed distribution 
function again satisfies \exref{eq:g}, but $\ephi$ is now an external field 
with prescribed statistical properties, entirely decoupled from $g$. 
This, of course, corresponds to setting $\alpha=0$ in \exref{eq:phi}. 

\section{Formalism: phase mixing and anti-phase-mixing}
\label{sec:formalism}

The spectral formalism for handling phase-space turbulence that we will use here was developed, 
for a different problem, by \citet{sch16} (see also \citealt{parker15}, \citealt{kanekar15}), 
but there are enough 
minor differences with this work to justify a detailed recapitulation. 
However, a reader already familiar with this material might save time 
by fast-forwarding to \eqref{eq:F} and then working her way backwards whenever 
anything appears unclear.  

\subsection{Hermite moments: waves and phase mixing}
\label{sec:fluid}

We will work in the Fourier--Hermite space, decomposing the perturbed distribution 
as follows
\beq
g(x,v) = \sum_k e^{ikx} \sum_m \frac{H_m(v)\FM(v)}{\sqrt{2^m m!}}\, g_{k,m},  
\quad
g_{k,m} = \int\frac{\rmd x}{2\pi L}\,e^{-ikx} 
\int\rmd v\, \frac{H_m(v)}{\sqrt{2^m m!}}\, g(x,v),  
\eeq
where $L$ is the system size. 
The Hermite polynomials 
\beq
H_m(v) = (-1)^m e^{v^2}\frac{\rmd^m}{\rmd v^m} e^{-v^2},\quad
\int\rmd v\, H_m(v) H_n(v) \FM(v) = 2^m m!\, \delta_{mn},
\eeq
form a convenient orthogonal basis for handling 1D perturbations to a Maxwellian. 
It is in anticipation of the Hermite decomposition that the Lenard--Bernstein 
collision operator \exref{eq:LB} was chosen, as the Hermite polynomials are its 
eigenfunctions: 
\beq
C[g_{k,m}] = - \nu m g_{k,m}. 
\eeq
To enforce momentum and energy conservation, we overrule this with $C[g_{k,1}]=0$ 
and $C[g_{k,2}]=0$.

Using the identities
\beq
v H_m = \frac{1}{2}\,H_{m+1} + m H_{m-1},\quad
\frac{\rmd H_m}{\rmd v} = 2m H_{m-1}, 
\eeq
denoting the particle density, flow velocity and temperature by 
\beq
n_k = g_{k,0},\quad \uu_k = \int\rmd v\,v g_{k}(v) = \frac{g_{k,1}}{\sqrt{2}},\quad
T_k = \sqrt{2}\,g_{k,2},  
\eeq
and noticing that \exref{eq:phi} then amounts to 
\beq
\ephi_k=\alpha_k n_k + \chi_k,
\label{eq:phik}
\eeq
we arrive at the following spectral representation of \exref{eq:g}: 
\begin{align}
\label{eq:g0}
&\frac{\dd n_k}{\dd t} + ik\uu_k = 0,\\
\label{eq:g1}
&\frac{\dd \uu_k}{\dd t} + ik\lt(\frac{T_k}{2} 
+ \frac{1+\alpha_k}{2}\,n_k\rt) 
+ \frac{1}{2}\sum_p ip \ephi_p n_{k-p} = -\frac{ik\chi_k}{2},\\
\label{eq:g2}
&\frac{\dd T_k}{\dd t} + ik\lt(\sqrt{3}\,g_{k,3} + 2\uu_k\rt) 
+ 2\sum_p ip \ephi_p \uu_{k-p} = 0,\\
\label{eq:gm}
&\frac{\dd g_{k,m}}{\dd t} 
+ ik\lt(\sqrt{\frac{m+1}{2}}\,g_{k,m+1} + \sqrt{\frac{m}{2}}\,g_{k,m-1}\rt)
+ \sqrt{\frac{m}{2}}\sum_p ip \ephi_p g_{k-p,m-1} = -\nu m g_{k,m},
\end{align}
the last equation describing all $m\ge3$.  

In the absence of sources, nonlinearities and heat fluxes ($g_{k,3}=0$), 
\exsdash{eq:g0}{eq:g2} describe 1D hydrodynamics of 
plasma waves (Langmuir waves for the electron model 
and ion-acoustic waves for the ion one). This becomes particularly 
obvious if we work in terms of the linear eigenfunctions
\beq
n^\pm_k = n_k \pm \frac{k}{\omega_k}\,\uu_k,\quad
\omega_k = k\sqrt{\frac{3+\alpha_k}{2}}, 
\label{eq:npm_def}
\eeq
denote $\theta_k = T_k - 2n_k$ (the non-adiabatic part of the temperature)
and recast \exsdash{eq:g0}{eq:g2} as follows
\begin{align}
\label{eq:npm}
&\frac{\dd n^\pm_k}{\dd t} \pm i\omega_k n^\pm_k 
= \mp \frac{ik}{2\omega_k}
\lt[k(\chi_k + \theta_k) 
+ \sum_p p\ephi_pn_{k-p}\rt],\\
&\frac{\dd\theta_k}{\dd t} = -ik\sqrt{3}\,g_{k,3} 
- 2\sum_pip\ephi_p\uu_{k-p}.
\label{eq:theta}
\end{align}
At any given $k$, the fluctuating fields $n^\pm_k$ oscillate at frequency $\omega_k$ 
(the Langmuir frequency or the ion sound frequency), with their energy injected 
by the forcing $\chi_k$. The nonlinear term, which, 
since $\ephi_p=\alpha_p n_p + \chi_p$, includes both self-interaction and advection 
by the external potential $\chi_p$, can, in general, transfer wave 
energy to different wave numbers, drain it or inject it. The term containing $\theta_k$ 
connects the wave dynamics to the entire hierarchy of higher Hermite moments, 
which evolve according to \exref{eq:gm}. 

In a linear system, the latter effect would give rise to Landau damping: 
the coupling of lower-order Hermite moments to higher-order ones 
that appears in the second term on the left-hand side of \exref{eq:gm} 
``phase-mixes'' perturbations to ever higher $m$'s, which represents 
emergence of ever finer structure in velocity space (at large $m$, 
the Hermite transform is effectively similar to a Fourier 
transform in $v$, with ``frequency'' $\sqrt{2m}$, so moments of order 
$m$ represent velocity-space structures with scale $\delta v \sim \pi/\sqrt{m}$).  
Eventually this 
activates collisions, however small their frequency $\nu$ might be, and the 
dynamics become irreversible. In the presence of nonlinearity, 
the situation is more complicated, with the last term on the 
left-hand side of \exref{eq:gm} causing a kind of advection of higher Hermite 
moments by the wave field~$\ephi_k$. This gives rise to filamentation 
of the distribution function not just in velocity but also in position 
space \citep{oneil65,manheimer71,dupree72}. 
The resulting coupling between different wave numbers can 
trigger plasma echos \citep{gould67}, or anti-phase-mixing, leading 
to cancellation, on average, of the Landau damping \citep{parker15,sch16,parker16}. 
It is with the latter phenomenon that we will be concerned in what follows, as we 
seek to characterise both spatial and velocity structure of the distribution 
function in terms of its $k$ and $m$ spectra. 

\subsection{Energy fluxes} 

We can define the energy spectrum of our waves to be 
\beq
W_k = \frac{3+\alpha_k}{2}\lt\la|n_k|^2\rt\ra + \lt\la|\uu_k|^2\rt\ra 
=  \frac{\omega_k^2}{2k^2} \lt\la|n_k^+|^2 + |n_k^-|^2\rt\ra. 
\eeq 
Using \exref{eq:npm}, we find that it evolves according to
\beq
\frac{\dd W_k}{\dd t} = k\,\Im\lt\la\chi_k\uu^*_k\rt\ra 
+ k\,\Im\lt\la\theta_k u_k^*\rt\ra
+ \sum_p p\,\Im\lt\la\ephi_p n_{k-p} \uu_k^*\rt\ra.
\label{eq:Wk}
\eeq
The first term on the right-hand side is the energy injection, 
the last term involves interactions between 
waves (it does not in general integrate 
to zero because waves can exchange energy, nonlinearly, with particles), 
whereas the second term is responsible for energy removal via phase mixing. 
We can see how this is picked up by higher Hermite moments if we define 
the Fourier-Hermite spectrum and Hermite flux\footnote{Note the extra factor 
of $1/2$ used here compared to the analogous quantities in \citet{sch16} and 
the typo (a missing minus sign) in the last expression for $\Gamma_m$ in their 
equation~(3.19).} 
\beq
C_{k,m} = \frac{1}{2}\lt\la|g_{k,m}|^2\rt\ra,\quad
\Gamma_{k,m} = k \sqrt{\frac{m+1}{2}}\,\Im\lt\la g_{k,m+1}^* g_{k,m}\rt\ra
\label{eq:CGdef}
\eeq  
and deduce from \exref{eq:gm} the evolution equation for the spectrum: 
\beq
\frac{\dd C_{k,m}}{\dd t} + \Gamma_{k,m} - \Gamma_{k,m-1} + 2\nu m C_{k,m} 
= \sqrt{\frac{m}{2}}\sum_p p\, \Im\lt\la\ephi_p g_{k-p,m-1} g_{k,m}^*\rt\ra.
\label{eq:Ckm}
\eeq
The phase-mixing term in \exref{eq:Wk} is 
\beq
k\,\Im\la\theta_k u_k^*\ra = k\lt[\Im\lt\la g_{k,2}g_{k,1}^*\rt\ra 
- 2\lt\la n_k\uu_k^*\rt\ra\rt] 
= -\Gamma_{k,1} + \frac{\dd}{\dd t}\lt\la|n_k|^2\rt\ra
\eeq
(the sloshing about of the fluctuation energy associated with the wave motion,
represented by the time derivative,  
averages out in the statistical steady state). The Hermite flux 
in \exref{eq:Ckm} passes energy along to higher $m$'s until the collision term 
is large enough to erase it. Our strategy will be to work out a universal 
form for $\Gamma_{k,m}$ in a turbulent plasma at high $m$.
Ideally, one would deduce from that what $\Gamma_{k,1}$ is, on average. 
In practice, we shall be able to predict that if one keeps a certain 
unspecified order-unity number $m$ of Hermite moments (``order-unity'' meaning 
finite and independent of collisionality, however small the latter is), 
the energy flux $\Gamma_{k,m}$
from/to these moments to/from the rest of phase space is zero in a certain 
``inertial'' range of wave numbers~$k$. 

\subsection{High-$m$ dynamics}
\label{sec:high_m}

Let us focus on the dynamics at $m\gg1$, deep in phase space (one might 
think of this as the ``inertial range'' of phase-space turbulence).~If 
\beq
p\ephi_p \ll k,\quad
\frac{\omega_k}{k} = \sqrt{\frac{3+\alpha_k}{2}} \ll \sqrt{m}, 
\label{eq:cond_cont}
\eeq 
then, to lowest order in $1/\sqrt{m}$, \exref{eq:gm} gives us simply  
\beq
g_{k,m+1} \approx - g_{k,m-1}.
\label{eq:lowestorder}
\eeq
This implies 
\beq
g_{k,m+1} \approx \pm i g_{k,m},
\label{eq:pm}
\eeq
i.e., $i^m g_{k,m}$ is either continuous or sign-alternating. 
When $k>0$, these two possibilities correspond to phase-mixing and anti-phase-mixing 
modes, respectively, and vice versa for $k<0$ \citep{kanekar15,parker15,sch16}. 
Let us separate these two cases explicitly. 

In view of \exref{eq:lowestorder}, the function 
\beq
\GG_{k,m} = \frac{i^m g_{k,m} + i^{m+1} g_{k,m+1}}{2} 
\label{eq:defG}
\eeq
will be approximately continuous in $m$: indeed, 
\beq
\GG_{k,m} - \GG_{k,m-1} = \frac{i^{m+1}}{2}\lt(g_{k,m+1} + g_{k,m-1}\rt) \approx 0. 
\eeq
Therefore, it is legitimate to treat $m$ as a continuous 
variable and approximate 
\beq
\GG_{k,m+1} \approx \GG_{k,m} + \frac{\dd \GG_{k,m}}{\dd m},\quad
\GG_{k,m-1} \approx \GG_{k,m} - \frac{\dd \GG_{k,m}}{\dd m},\quad
\text{etc.,}
\eeq
treating the derivative terms as small. Using this approximation, 
we can cast \exref{eq:gm} in the following approximate form, 
valid to lowest order in $1/\sqrt{m}$, 
\beq
\frac{\dd \GG_{k,m}}{\dd t} + \sqrt{2}\,k\, m^{1/4}\frac{\dd}{\dd m} m^{1/4} \GG_{k,m} 
+ \nu m \GG_{k,m} = \sqrt{\frac{m}{2}}\sum_p p \ephi_p \GG_{k-p,m}. 
\label{eq:Gm}
\eeq 
Finally, if we define 
\beq
\tf_k(s) = m^{1/4} \GG_{k,m}, \quad s= \sqrt{m}, 
\eeq
\eqref{eq:Gm} becomes 
\beq
\frac{\dd\tf_k}{\dd t} + \frac{k}{\sqrt{2}}\frac{\dd\tf_k}{\dd s}  
+ \nu s^2 \tf_k = \frac{s}{\sqrt{2}}\sum_p p \ephi_p \tf_{k-p}. 
\label{eq:tf}
\eeq 
This is very similar to equation (3.12) of \citet{sch16} and we have kept 
their notation for a faithful reader's convenience.\footnote{\citet{sch16} 
constructed the function $\tf_k$ by 
first separating $g_{k,m}$ into phase-mixing and anti-phase-mixing 
modes, $g_{k,m}^\pm$, then splicing those together into $\tf_k$, 
with positive $k$'s corresponding to $g^+_{k,m}$ and negative $k$'s 
to $g^-_{k,m}$. The equivalence of this approach to the shorter route 
via $\GG_{k,m}$ defined in \exref{eq:defG} was pointed out to us by W.~Dorland.
Note that \exref{eq:tf} is almost exactly the equation that we would have 
obtained by Fourier transforming the kinetic equation 
\exref{eq:g} in both $x$ and $v$, 
with $\sqrt{2}\,s$ in the role of the dual variable to $v$  
\citep[cf.][]{knorr77}, but we prefer the Hermite-transform approach.} 
It is manifest in \eqref{eq:tf} that when $k>0$, $\tf_k$ propagates to 
higher $s$ (phase-mixes) and when $k<0$, it propagates to lower $s$ 
(anti-phase-mixes) and that the coupling between wave numbers in the 
nonlinear term can turn phase-mixing perturbations into anti-phase-mixing ones 
and vice versa. 

The distribution function itself can be reconstructed from 
$\tf_k$, or, equivalently, from $\GG_{k,m}$, as follows\footnote{Note that, whereas 
$g_{k,m}$, being a Fourier transform of a real function, 
must satisfy $g^*_{-k,m} = g_{k,m}$, neither $\GG_{k,m}$ nor $\tf_k$ are subject
to any such constraint and indeed one can show that $\tf^*_{-k} = \tf_k$ 
only in the absence of phase mixing \citep{sch16}.} 
\beq
g_{k,m} = (-i)^m \GG_{k,m} + i^m \GG^*_{-k,m}.
\label{eq:gm_rec}
\eeq
The fact that, for any given $k$, both $\GG_{k,m}$ and $\GG_{-k,m}$ are necessary 
to reconstruct $g_{k,m}$ reflects the presence of both phase-mixing and 
anti-phase-mixing modes in any distribution function. Maintaining a solution 
of \exref{eq:tf} with no anti-phase-mixing, viz., $\tf_k = 0$ for all $k<0$, 
is clearly only possible in the absence of the nonlinearity.  

Finally, if we define 
\beq
F_k = \bigl\la|\tf_k|^2\bigr\ra = \sqrt{m} \lt\la |\GG_{k,m}|^2 \rt\ra,
\eeq
both the Fourier-Hermite spectrum $C_{k,m}$ and the Hermite flux $\Gamma_{k,m}$, 
defined in \exref{eq:CGdef}, can be reconstructed from $F_k$: 
\begin{align}
\label{eq:C_F}
C_{k,m} & \approx \frac{C_{k,m} + C_{k,m+1}}{2} 
= \frac{\lt\la|\GG_{k,m}|^2 + |\GG_{-k,m}|^2\rt\ra}{2}
= \frac{F_k + F_{-k}}{2\sqrt{m}},\\
\Gamma_{k,m} & = k\sqrt{\frac{m+1}{2}}\lt\la|\GG_{k,m}|^2-|\GG_{-k,m}|^2\rt\ra
\approx \frac{k}{\sqrt{2}}\lt(F_k - F_{-k}\rt). 
\label{eq:Gamma_F}
\end{align}
The derivation of these relations relies on \exref{eq:gm_rec} and on noticing also that 
\beq
ig_{k,m+1} = (-i)^m \GG_{k,m} - i^m \GG_{-k,m}^*. 
\eeq
Thus, the Fourier-Hermite spectrum is the average of the spectra 
of the phase-mixing and anti-phase-mixing modes and the Hermite 
flux is their difference. It remains to solve for $F_k$, which, 
using \exref{eq:tf}, is immediately found to satisfy 
\beq
\frac{\dd F_k}{\dd t} + \frac{k}{\sqrt{2}}\frac{\dd F_k}{\dd s}  
+ 2 \nu s^2 F_k = s \sqrt{2}\, \Re \sum_p p \bigl\la \ephi_p \tf_{k-p} \tf_k^*\bigr\ra. 
\label{eq:F}
\eeq 
This equation, which is an approximate continuous version of \exref{eq:Ckm}, 
is not closed and so we will need a plausible method for handling its right-hand side.

\section{Method: Kraichnan--Batchelor limit} 
\label{sec:method}

\subsection{Kraichnan--Kazantsev model}
\label{sec:KK}

We shall be brutal and obtain a closure for the right-hand side of \eqref{eq:F} 
by modelling $\ephi_p$ as a random Gaussian white-noise 
(short-time-correlated) field,\footnote{The potential usefulness of this model for the 
Vlasov equation appears to have been first recognised in an elegant paper by \citet{cook78}, 
who derived some relevant equations, 
discussed their relationship to various other approaches that were being 
tried in the 1960s and 70s, and promised solutions, 
but did not, it seems, follow up. We note that \citet{orszag67} appear to have been 
the first to pose phase-space correlations of $\df$ in a Vlasov plasma with a stochastic 
electric field as a worthwhile problem, substantially influencing the field, without, 
however, providing solutions.} 
\beq
\lt\la\ephi_p(t)\ephi_{p'}(t')\rt\ra = 2 \kap_p \delta_{p,-p'}\delta(t-t'). 
\label{eq:KK}
\eeq
This assumption, pioneered by \citet{kraichnan68} (for the passive-scalar problem) 
and \citet{kazantsev68} (for the turbulent-dynamo problem) 
is of course quantitatively wrong, but there is a long and encouraging history in 
fluid dynamics and MHD of the resulting closure leading to results that are basically 
correct \citep[e.g.,][]{kraichnan74,kraichnan94,zeldovich90,krommes97clumps,falkovich01,boldyrev04,sch04dynamo,sch04scalar,sch07dynamo,bhat15}. In the present context, what we are doing formally amounts 
to ignoring the contribution of the density $n_k=g_{k,0}$ to $\ephi_k$ in \exref{eq:phik}
and stipulating the statistics \exref{eq:KK} for the external forcing $\chi_k$. 
Physically, we are assuming that the advecting stochastic electric field can be treated as 
statistically independent of the phase-space structure of the distribution function. 
A reader unconvinced that this can ever be a valid approximation 
for any aspect of the Vlasov problem with a self-consistent electric field, might find 
comfort in considering the calculations that follow to apply solely to the 
stochastic-acceleration problem, where $\ephi=\chi$ (\secref{sec:stoch_acc}). 

For a Gaussian field, by the theorem of \citet{furutsu63} and \citet{novikov65}, 
\beq
\bigl\la \ephi_p \tf_{k-p} \tf_k^*\bigr\ra(t) 
= \int^t\rmd t'\sum_{p'}\lt\la\ephi_p(t)\ephi_{p'}(t')\rt\ra 
\lt\la\frac{\delta[\tf_{k-p}(t)\tf_k^*(t)]}{\delta\ephi_{p'}(t')}\rt\ra.
\label{eq:FN}
\eeq
Using \exref{eq:tf} to write an evolution equation for $\tf_{k-p}(t)\tf_k^*(t)$ 
and then formally integrating it over time, we find 
\begin{align}
\nonumber
\tf_{k-p}(t)\tf_k^*(t) &= \int^t\rmd t''\Biggl\{-\frac{k}{\sqrt{2}}\tf_{k-p}\frac{\dd\tf_k^*}{\dd s}
-\frac{k-p}{\sqrt{2}}\tf_k^*\frac{\tf_{k-p}}{\dd s}
- 2\nu s^2\tf_{k-p}\tf_k^*\Biggr.\\
&\Biggl.\qquad + \frac{s}{\sqrt{2}}\sum_{p''}p''\lt[\ephi_{p''}\tf_{k-p-p''}\tf_k^* 
+ \ephi_{-p''}\tf_{k-p''}^*\tf_{k-p}\rt]
\Biggr\}(t'').
\end{align}
Therefore, its functional derivative is 
\beq
\lt\la\frac{\delta[\tf_{k-p}(t)\tf_k^*(t)]}{\delta\ephi_{p'}(t')}\rt\ra = 
\frac{s}{\sqrt{2}}\, p' \lt[\lt\la \tf_{k-p-p'}(t')\tf_k^*(t')\rt\ra 
- \lt\la \tf_{k+p'}^*(t')\tf_{k-p}(t')\rt\ra + \dots\rt] H(t-t'),
\label{eq:ddphi}
\eeq
where $H(t-t')$ is the Heaviside function, expressing the fact that,  
by causality, $\tf(t)$ cannot depend on $\ephi(t')$ at a future time $t'>t$, 
and ``$\dots$" stands for terms that vanish when $t'=t$.   
Substituting \exref{eq:KK} and \exref{eq:ddphi} into \exref{eq:FN} gives
\beq
\bigl\la \ephi_p \tf_{k-p} \tf_k^*\bigr\ra = 
-\frac{s}{\sqrt{2}}\,p\kap_p \lt(F_k - F_{k-p}\rt).
\eeq 
Finally, using this in \exref{eq:F}, we get 
\beq
\frac{\dd F_k}{\dd t} + \frac{k}{\sqrt{2}}\frac{\dd F_k}{\dd s}  
+ 2 \nu s^2 F_k = s^2\sum_p p^2 \kap_p \lt(F_{k-p} - F_k\rt). 
\label{eq:Fclosed}
\eeq
The $F_k$ term on the right-hand side is an additional, ``turbulent'' collisionality 
(turbulent diffusion in velocity space); the $F_{k-p}$ term is the mode-coupling 
term responsible for moving energy around and for converting phase-mixing modes 
into anti-phase-mixing ones or vice versa. 

In steady state, $\dd F_k/\dd t = 0$ and \exref{eq:Fclosed} can be recast 
in an even simpler form: dividing through by $2s^2$, we arrive at
\beq
k\frac{\dd F_k}{\dd\tau} + \nu F_k = \frac{1}{2}\sum_p p^2 \kap_p \lt(F_{k-p} - F_k\rt),
\quad \tau \equiv \frac{(\sqrt{2}\,s)^3}{3} = \frac{(2m)^{3/2}}{3}.
\label{eq:Fstst}
\eeq

\subsection{Energy budget and collisions}
\label{sec:budget}

It is an important property of the Kraichnan--Kazantsev model applied to our 
problem that, in \exref{eq:Fclosed}, the nonlinear 
interactions disappear under summation over all $k$ and so 
the total ``energy'' of the $\tf$ field has a conservation law: 
\beq
\frac{\rmd}{\rmd t}\int_{\smin}^\infty\rmd s\sum_k F_k = \frac{1}{\sqrt{2}} 
\sum_k k F_k(\smin) - 2\nu \int_{\smin}^\infty\rmd s\, s^2 \sum_k F_k,
\label{eq:budget}
\eeq 
where $\smin$ is some suitably chosen lower cutoff and the energy balance is between 
the flux through that cutoff (from or towards the waves at low $m$) and 
collisional dissipation. Restating this in the steady state and with the $\tau$ 
variable \exref{eq:Fstst}, 
\beq
\int_{-\infty}^{+\infty}\rmd k\,k F_k(\tmin) = \nu\int_{\tmin}^\infty\rmd\tau 
\int_{-\infty}^{+\infty}\rmd k F_k(\tau).
\label{eq:budget_stst}
\eeq
This energy balance admits two distinct physical scenarios. 

One is essentially similar 
to what happens in the absence of nonlinearity ($\kap_p=0$): in the limit of small $\nu$, 
the spectrum $F_k$ is independent of $\tau$ (or of $s$, or of $m$), giving us the shallow 
$C_{k,m}\propto 1/\sqrt{m}$ slope associated with a Landau-damped 
solution \citep{zocco11,kanekar15}. Collisions become important 
at $\tau\sim \nu^{-1}$ [see \exref{eq:Fstst}] and so the dissipation 
term in the right-hand side of \exref{eq:budget_stst} is finite and independent 
of collisionality as $\nu\to+0$. This in turn implies that the integral in left-hand 
side of \exref{eq:budget_stst} must be finite and non-zero (in the linear regime, 
$F_{k<0}=0$, so the integral is always positive and will be 
finite as long as the wave-number spectrum decays fast enough). 

The second scenario arises from any Hermite spectral slope that makes 
$F_k(\tau)$ decay with $\tau$. Then the collisional dissipation vanishes 
as $\nu\to+0$, i.e., the limit of vanishing collisionality is non-singular 
in this sense, giving us license simply to set $\nu=0$ in \exref{eq:Fstst} 
and expect to find a legitimate solution. The solution is 
indeed a legitimate steady-state solution if the integral in the left-hand side 
of \exref{eq:budget_stst} vanishes for it, i.e., if the overall energy flux 
into Hermite space is zero: 
\beq
\int_{-\infty}^{+\infty}\rmd k\,k F_k(\tmin) = 
\int_{0}^{+\infty}\rmd k\,k \lt[F_k(\tmin) - F_{-k}(\tmin)\rt] = 
\int_{0}^{+\infty}\rmd k\sqrt{2}\, \Gamma_{k,\mmin} = 0, 
\label{eq:zeroflux_tot}
\eeq 
although there is no {\em a priori} requirement that the Hermite 
flux must vanish at every $k$. 
We shall see that this is exactly the state that emerges in the nonlinear regime. 

\subsection{Batchelor limit}
\label{sec:Batchelor}

While \exref{eq:Fstst} is a closed and compact equation, it is an integral 
one and not necessarily easily amenable to analytical solution. 
We are going to make a further simplification by assuming that 
$p^2\kap_p$ decays sufficiently steeply with $p$ 
that it is meaningful to consider $F_k$ at $|k|\gg p$, an approach pioneered 
by \citet{batchelor59} in the context of passive-scalar mixing 
\citep[with the more quantitative theory due to][]{kraichnan74}. 
We can then expand under the wave-number sum in \exref{eq:Fstst}:
\beq
F_{k-p} - F_k \approx - p\frac{\dd F_k}{\dd k} + \frac{1}{2}\,p^2\frac{\dd^2 F_k}{\dd k^2}.
\label{eq:Batchelor_limit}
\eeq  
Having noticed that the first term vanishes under summation because 
it is odd in $p$ while $\kap_p=\kap_{-p}$, we obtain a rather simple differential equation: 
\beq
k\frac{\dd F_k}{\dd\tau} + \nu F_k = \gamma \frac{\dd^2 F_k}{\dd k^2},\quad
\gamma \equiv \frac{1}{4}\sum_p p^4\kap_p. 
\label{eq:Floc}
\eeq  
The wave-number-diffusion rate $\gamma$ can easily be scaled out. 

It turns out (see \apref{ap:Zslns}) that $p^2\kap_p$ must decay 
more steeply than $p^{-2}$ at the very least, in order for this 
approximation to make sense, although it would have to be steeper 
than $p^{-3}$ in order for the $p$ integral that determines $\gamma$ 
in \exref{eq:Floc} to converge without the need 
for a high-$p$ cutoff. In what follows, we shall effectively 
assume the advecting electric field to be single-scale, with $\kap_p$ 
concentrated around some characteristic wave number~$p$. 

Equation \exref{eq:Floc} makes it clear how the nonlinearity causes anti-phase-mixing. 
At $k>0$, the steady-state \eqref{eq:Floc} can be thought of as a diffusion 
equation in $k$ (or, to be precise, in $|k|^{3/2}$; see \secref{sec:plan}),  
with $\tau$ playing the role of time. At $k<0$, this ``time'' reverses, i.e., 
diffusion turns into antidiffusion. Whatever energy resides at any given 
$k>0$ and low $\tau$ will, as $\tau$ increases, spread over the $k$ space. 
Some of it can spread towards $k=0$, where it crosses into the $k<0$ 
territory (the rules of this crossing are established in \secref{sec:bc}) 
and diffuses back towards large (negative) $k$ and low $\tau$. 
This creates an anti-phase-mixing energy flux, which can (and will) cancel 
the phase-mixing flux. Collisions limit the values of $\tau$ (and of $k$) 
available to these phase-space flows. We shall discuss 
their role more quantitatively in \secref{sec:colls}, after we have the exact 
collisionless solution in hand.  

\subsection{Boundary conditions: continuity in $k$ space} 
\label{sec:bc}

We would like to be able to treat the solution 
of \exref{eq:Fstst} in the region of low $|k|\sim p$, where the Batchelor 
approximation is not valid, as a continuous extension 
of the solution of \exref{eq:Floc}. In other words, we wish to prove that 
we can simply solve \exref{eq:Floc} with the boundary conditions
\beq
F_{k\to+0} = F_{k\to-0},\quad
\lt.\frac{\dd F_k}{\dd k}\rt|_{k\to + 0} = \lt.\frac{\dd F_k}{\dd k}\rt|_{k\to - 0},
\label{eq:zero_cont}
\eeq 
where $k\to\pm0$ really means $k\to\pm p$. 

The continuity of $F_k$ across $k=0$ and all the way to $|k|\gg p$ 
(i.e., to the wave numbers where the Batchelor approximation holds)
can be inferred from \exref{eq:Fstst} as follows. Let us ignore collisions and 
consider sufficiently small $k$ (and/or sufficiently large $\tau$) 
so that the phase-mixing term is small compared to the nonlinear 
term: from \exref{eq:Floc}, this is true for $|k|\ll(\gamma\tau)^{1/3}$, 
which can be satisfied already at $|k|\gg p$, at least for $\tau\gg p^3/\gamma$. 
Then \exref{eq:Fstst} reduces to
\beq
\sum_p p^2\kap_p F_{k-p} = \lt(\sum_p p^2\kap_p\rt) F_k.
\label{eq:nlin_zero}
\eeq
If we denote $p^2\kap_p = K_p$, assume that the width of this function is 
smaller than the range of $k$ in which \exref{eq:nlin_zero} is valid 
(unlike in \apref{ap:Zslns}, where the validity of this approach is probed), 
turn sums in \exref{eq:nlin_zero} into integrals and 
Fourier-transform \exref{eq:nlin_zero}, denoting the dual variable by $x$, 
and endowing the transformed functions with hats, we get 
\beq
\hat K(x) \hat F(x) = \hat K(0) \hat F(x). 
\eeq 
The solution of this equation is $\hat F(x)\propto\delta(x)$. 
Therefore, $F_k$ is independent of $k$ in a range of $k$ surrounding $0$, 
with characteristic width $\sim p$, the typical wavenumber of the advecting 
field~$\ephi$.

The second of the relations \exref{eq:zero_cont} 
(the continuity of the derivative) follows from 
the requirement that there can be no total energy flux in or out of $k=0$ 
(because $C_{k,m}$ must be even in $k$ by the reality condition). 
Therefore, from \exref{eq:C_F},  
\beq
\lt.\frac{\dd}{\dd k}\lt(F_k + F_{-k}\rt)\rt|_{k\to+0}=0
\quad\Rightarrow\quad 
\lt.\frac{\dd F_k}{\dd k}\rt|_{k\to+0} 
= - \lt.\frac{\dd F_{-k}}{\dd k}\rt|_{k\to+0} 
= \lt.\frac{\dd F_k}{\dd k}\rt|_{k\to-0}. 
\eeq

\section{Solution: universal self-similar phase-space spectrum}
\label{sec:solution}

Although we are about to entertain ourselves and the reader with an exact solution 
of \exref{eq:Floc}, the morphology of this solution is, in fact, not hard to grasp already 
by a cursory examination of the equation. We will discuss it {\em post hoc}, in 
\secref{sec:colls}. A reader with no time for mathematical niceties 
can start from there and leaf back as necessary. 

\subsection{Plan of solution}
\label{sec:plan}

If, as we promised in \secref{sec:budget}, 
we are going to discover that phase mixing is substantially or fully suppressed, 
we must be able find such a solution from \exref{eq:Floc} with $\nu=0$. 
Note that the $\tau$ variable can now be shifted arbitrarily and so we can choose 
$\tau=0$ to correspond to any true value of $m$---so let 
$\tau-\tmin \to \tau$, where $\tmin$ is the lower cutoff introduced 
in \secref{sec:budget}. Physically, the limit $\tau\to0$ corresponds to going back 
from the depths of phase space to low $m$'s, where the approximations that led to 
\exref{eq:tf} break down. 

With these further simplifications, \exref{eq:Floc} turns into 
a type of diffusion equation at $k>0$ and antidiffusion at $k<0$. 
Namely, letting 
\beq
F_k = \lt\{\begin{array}{l} 
F^+(\xx),\quad k>0\\
F^-(\xx),\quad k<0
 \end{array}\rt.,\quad
\xx \equiv \frac{2}{3\sqrt{\gamma}}|k|^{3/2},
\eeq
we find that $F^\pm$ satisfies
\beq
\pm \frac{\dd F^\pm}{\dd\tau} = \frac{1}{\xx^{1/3}}\frac{\dd}{\dd \xx} 
\xx^{1/3}\frac{\dd F^\pm}{\dd \xx}.
\label{eq:Fpm}
\eeq
The ``$+$'' version of this equation 
belongs to a class studied exhaustively by \citet{sutton43}, who derived its 
Green's functions for all standard initial and boundary-value problems. 
Armed with these, we are going to construct the full solution in the following way.

(i) First postulate an ``initial'' condition and a boundary condition for $F^+$ 
and find the Green's-function solution for $F^+(\tau,\xx)$: 
\beq
\lt.\begin{array}{l}
F^+(\tau\to0,\xx) = F_0^+(\xx),\\
F^+(\tau,\xx\to0) = Y(\tau)
\end{array}
\rt\}
\quad\Rightarrow\quad F^+(\tau,\xx),
\label{eq:icbc_plus}
\eeq
where $F_0^+(\xx)$ and $Y(\tau)$ are some unknown functions.
Obviously, as we are not interested in spectra that blow up at 
small scales, $F^+$ must vanish at $\xi\to\infty$ 
and the same will be required of $F^-$. 

(ii) Since we must have continuity, $F^+(\tau,\xx\to0) = F^-(\tau,\xx\to0)$ 
[see \exref{eq:zero_cont}], 
$F^-(\tau,\xx)$ is found as a Green's-function solution with an unknown 
value $Y(\tau)$ on the boundary.
Since the equation for $F^-$ is an antidiffusion equation, the ``initial'' condition 
for it must be set at large $\tau$ and since there can be no energy at $\tau\to\infty$, 
this ``initial'' condition is zero. Formally, this can be implemented by choosing 
some cutoff $\tmax$, requiring the function to vanish there, solving, and then 
taking $\tmax\to\infty$. The validity of this operation will be confirmed by the 
finiteness of the result. Thus,
\beq
\lt.\begin{array}{l}
F^-(\tau,\xx\to0) = Y(\tau),\\
F^-(\tau\to\tmax,\xx) = 0 
\end{array}
\rt\}
\quad\Rightarrow\quad F^-(\tau,\xx),\quad
\tmax\to\infty.
\label{eq:icbc_minus}
\eeq
Operationally, the solution can be accomplished by 
changing variables to $\tau' = \tmax - \tau$, so the antidiffusion equation 
turns into a diffusion one. Physically, $\tmax\sim 1/\nu$.

(iii) The unknown function $Y(\tau)$ is now determined by the continuity of the 
energy flux across $k=0$ [see \exref{eq:zero_cont}]. 
Since $\dd F_k/\dd k = \pm\xx^{1/3}\dd F^{\pm}/\dd\xx$, 
\beq
\lt.\xx^{1/3}\frac{\dd F^+}{\dd\xx}\rt|_{\xi\to0} = 
- \lt.\xx^{1/3}\frac{\dd F^-}{\dd\xx}\rt|_{\xi\to0} 
\quad
\Rightarrow
\quad
Y(\tau). 
\label{eq:flux_cont}
\eeq
A key physical constraint is that $Y(\tau)$ should be a decreasing function,  
otherwise the assumption that the collisional dissipation is inessential 
would have to be abandoned. 

(iv) At this point, we are in possession of the full solution, subject to the 
unknown function $F^+_0(\xx)$. We may now use this solution to determine 
\beq
F^-_0(\xx) = F^-(\tau\to0,\xx). 
\eeq
The net Hermite flux \exref{eq:Gamma_F} at $\tau\to0$, i.e., from low Hermite moments 
to high ones, is proportional to $F^+_0-F^-_0$. We will show that there is a solution 
for which 
\beq
F^+_0-F^-_0 = 0
\eeq
and that this solution is the only physically sound one. Thus, the outcome 
of this procedure will be a universal structure of the Fourier--Hermite spectrum 
$F_k(\tau)$ in steady state. An impatient reader can skip what follows 
to find this spectrum in \secref{sec:ssim} (xe will also find a shorter, 
more elementary, if perhaps less general, route to this solution in \apref{ap:ssim}).

\subsection{Green's function solution}

\subsubsection{The ``$+$'' solution}

The solution of the ``$+$'' equation \exref{eq:Fpm} satisfying the initial 
and boundary conditions \exref{eq:icbc_plus} is 
\begin{align}
\nonumber
F^+(\tau,\xx) &= \frac{\xx^{1/3}e^{-\xx^2/4\tau}}{2\tau} \int_0^\infty\rmd\yy\,\yy^{2/3} 
e^{-\yy^2/4\tau} I_{1/3}\!\lt(\frac{\xx\yy}{2\tau}\rt) F_0^+(\yy)\\
&\qquad + \frac{\lt(\frac{1}{2}\xx\rt)^{2/3}}{\Gamma\!\lt(\frac{1}{3}\rt)}\int_0^\tau\rmd\sigma\,
\frac{e^{-\xx^2/4(\tau-\sigma)}}{(\tau-\sigma)^{4/3}}\,Y(\sigma),  
\label{eq:Fp_sln}
\end{align}
where $I_{1/3}$ is a modified Bessel function of the first kind. 
The first term in \exref{eq:Fp_sln} 
is responsible for satisfying the initial condition and is zero 
at $\xi\to0$, the second term is equal to zero at $\tau\to0$ and to $Y(\tau)$ at $\xi\to0$. 

The associated energy flux at $\xx\to0$ is 
\beq
\lt.\xx^{1/3}\frac{\dd F^+}{\dd\xx}\rt|_{\xx\to0}(\tau) 
= \frac{2^{1/3}}{\Gamma\!\lt(\frac{1}{3}\rt)}
\lt[\frac{1}{\tau^{1/3}}\int_0^\infty\rmd z\,e^{-z} F_0^+(2\sqrt{\tau z})
- \frac{\rmd}{\rmd\tau}\int_0^\tau\rmd\sigma\,\frac{Y(\sigma)}{(\tau-\sigma)^{1/3}}\rt].
\label{eq:flux_plus}
\eeq
The first term is obtained by expanding 
\beq
I_{1/3}\!\lt(\frac{\xx\yy}{2\tau}\rt) = \frac{1}{\Gamma\!\lt(\frac{4}{3}\rt)} 
\lt(\frac{\xx\yy}{4\tau}\rt)^{1/3} + \dots, 
\eeq
under the integral, mopping up powers of $\xx$, then taking $\xx\to0$, and, 
finally, changing the integration variable to $z = y^2/4\tau$. 
The second term in \exref{eq:flux_plus} takes some work---the derivation  
can be found in \citet{sutton43} (his equation 7.4).\footnote{The 
flux has to be manipulated into this form because simply taking $\xx=0$ in the 
second integral in \exref{eq:flux_plus} leads to a potentially divergent 
integral. The idea of the derivation is first to replace 
$Y(\sigma) = [Y(\sigma) - Y(\tau)] + Y(\tau)$ under the integral, 
do the integral multiplying $Y(\tau)$ exactly before taking $\xx\to0$, 
whereas in the integral involving $Y(\sigma)-Y(\tau)$ ensure convergence 
in the limit $\xx\to0$ by assuming sufficient regularity of the function $Y$. 
A few integrations by parts later, \eqref{eq:flux_plus} results.} 

\subsubsection{The ``$-$'' solution}

To obtain the solution of the ``$-$'' equation \exref{eq:Fpm}, let $\tau'= \tmax-\tau$
and solve 
\beq
\frac{\dd F^\pm}{\dd\tau'} = \frac{1}{\xx^{1/3}}\frac{\dd}{\dd \xx} 
\xx^{1/3}\frac{\dd F^\pm}{\dd \xx}
\eeq
subject to the boundary and initial conditions \exref{eq:icbc_minus}, which 
become, with the new variable,  
\beq
\begin{array}{l}
F^-(\tau',\xx\to0) = Y(\tmax - \tau'),\\
F^-(\tau'\to0,\xx) = 0. 
\end{array}
\eeq
The solution is the same as \exref{eq:Fp_sln}, but with $F_0^+$ replaced by $0$, 
$\tau$ by $\tau'$ and $Y(\sigma)$ by $Y(\tmax - \sigma)$. 
Changing the integration variable $\tmax - \sigma \to \sigma$ 
and restoring $\tau'= \tmax-\tau$, we get 
\beq
F^-(\tau,\xx) = \frac{\lt(\frac{1}{2}\xx\rt)^{2/3}}{\Gamma\!\lt(\frac{1}{3}\rt)}
\int_\tau^{\tmax}\rmd\sigma\,
\frac{e^{-\xx^2/4(\sigma-\tau)}}{(\sigma-\tau)^{4/3}}\,Y(\sigma).  
\label{eq:Fm_sln}
\eeq
Note that taking $\tmax\to\infty$ produces no anomalies, assuming $Y(\sigma)$ 
does not grow (which would not be physical anyway as even with linear Landau 
damping, $Y=\const$). 

The flux associated with this solution at $\xi\to 0$ is found from $F^-(\tau',\xx)$
by the same procedure as the second term in \exref{eq:flux_plus}, followed by the 
same changes of variables as described above. The result is 
\beq
\lt.\xx^{1/3}\frac{\dd F^-}{\dd\xx}\rt|_{\xx\to0}(\tau) 
= \frac{2^{1/3}}{\Gamma\!\lt(\frac{1}{3}\rt)}
\frac{\rmd}{\rmd\tau}\int_\tau^{\tmax}\rmd\sigma\,\frac{Y(\sigma)}{(\sigma-\tau)^{1/3}}.
\label{eq:flux_minus}
\eeq
Note that we should not rush into taking $\tmax\to\infty$ here before we take the 
derivative as the integral may well (and indeed will) prove divergent.

\subsubsection{Continuity of energy flux}

Using \exref{eq:flux_plus} and \exref{eq:flux_minus}, the condition \exref{eq:flux_cont} 
becomes 
\beq
\frac{\rmd}{\rmd\tau}\lt[
\int_0^\tau\rmd\sigma\,\frac{Y(\sigma)}{(\tau-\sigma)^{1/3}}
- \int_\tau^{\tmax}\rmd\sigma\,\frac{Y(\sigma)}{(\sigma-\tau)^{1/3}}
\rt]
=\frac{1}{\tau^{1/3}}\int_0^\infty\rmd z\,e^{-z} F_0^+(2\sqrt{\tau z}).
\label{eq:flux_cont_sln}
\eeq
This is an integral equation for $Y(\tau)$ in terms of the unknown function 
$F^+_0$ (or vice versa), but solving it directly is a thankless pursuit. 
We will cut to the chase and seek a particular solution for which 
the Hermite flux at $\tau\to0$ vanishes. 

\subsection{Zero-flux solution}
\label{sec:zeroflux}

From \exref{eq:Fm_sln}, we can deduce the spectrum of anti-phase-mixing 
modes at $\tau\to0$: 
\beq
F^-_0(\xx) = \frac{\lt(\frac{1}{2}\xx\rt)^{2/3}}{\Gamma\!\lt(\frac{1}{3}\rt)}
\int_\tau^{\tmax}\rmd\sigma\,
\frac{e^{-\xx^2/4\sigma}}{\sigma^{4/3}}\,Y(\sigma).  
\label{eq:Fm0}
\eeq
Let us explicitly look for such a solution that $F^-_0(\xx) = F^+_0(\xx)$. 
We may then substitute the expression \exref{eq:Fm0} for $F^+_0$ in \exref{eq:flux_cont_sln} 
and thus obtain an equation for what $Y(\tau)$ would have to be 
in order for the zero-flux solution to be realised. If this $Y(\tau)$ is physically 
legitimate (decays with $\tau$), we can put it back into \exref{eq:Fm0} 
to determine $F^\pm_0(\xx)$ and then use that and $Y(\tau)$ in 
\exref{eq:Fp_sln} and \exref{eq:Fm_sln} to determine the Fourier--Hermite 
spectrum everywhere. 

The integral in the right-hand side of \exref{eq:flux_cont_sln} with 
\exref{eq:Fm0} for $F^+_0$ can easily be done after switching the order of 
$z$ and $\sigma$ integrations and changing the integration variable $z$
to $\zeta = z(1+\tau/\sigma)$. As a result, \exref{eq:flux_cont_sln} becomes 
\beq
\frac{\rmd}{\rmd\tau}\lt[
\int_0^\tau\rmd\sigma\,\frac{Y(\sigma)}{(\tau-\sigma)^{1/3}}
- \int_\tau^{\tmax}\rmd\sigma\,\frac{Y(\sigma)}{(\sigma-\tau)^{1/3}}
\rt]
=\frac{1}{3}\int_0^\infty\rmd\sigma\,\frac{Y(\sigma)}{(\tau + \sigma)^{4/3}}.
\label{eq:flux_cont_zeroflux}
\eeq

The solution to \exref{eq:flux_cont_zeroflux} is, up to a multiplicative constant,
\beq
Y(\tau) = \frac{1}{\tau^{2/3}},
\label{eq:Y}
\eeq
which we will presently show by direct substitution. In \apref{ap:inevitability}, we  
show that this is indeed the only sensible solution. In \apref{ap:ssim}, this scaling 
with $\tau$ emerges via a much simpler (but less rigorous) 
argument arising from seeking a self-similar solution to our problem. 
 
With \exref{eq:Y} for $Y$, all integrals in \exref{eq:flux_cont_zeroflux} 
are standard tabulated ones: 
the integral in the right-hand side of \exref{eq:flux_cont_zeroflux} is 
(after rescaling the integration variable $\sigma/\tau\to\sigma$)
\beq
\text{r.h.s. of \exref{eq:flux_cont_zeroflux}} 
= \frac{1}{3\tau}\int_0^\infty\frac{\rmd\sigma}{\sigma^{2/3}(1+\sigma)^{4/3}}
= \frac{1}{\tau}, 
\label{eq:rhs}
\eeq
the first integral in the left-hand side is a constant (because $\tau$ can scaled out)
and the second one is dominated by the upper limit, so it is $\approx \ln (\tmax/\tau)$ 
as $\tmax\to\infty$. It then follows immediately that their derivative in 
the left-hand side of \exref{eq:flux_cont_zeroflux} is
\beq
\text{l.h.s. of \exref{eq:flux_cont_zeroflux}} \approx \frac{\rmd}{\rmd\tau}
\lt(\const - \ln\frac{\tmax}{\tau}\rt) = \frac{1}{\tau}
\quad\text{as}\quad \tmax \to \infty. 
\label{eq:lhs}
\eeq
This is equal to \exref{eq:rhs} and so \exref{eq:Y} is indeed a good solution. 

For future reference, these results imply that the energy flux through 
$\xx=0$ associated with this solution is 
\beq
\lt.\frac{\dd F_k}{\dd k}\rt|_{k\to0} = 
\pm\lt(\frac{3}{2\gamma}\rt)^{1/3}\lt.\xx^{1/3}\frac{\dd F^\pm}{\dd\xx}\rt|_{\xx\to0}(\tau) 
= \frac{3^{1/3}}{\gamma^{1/3}\Gamma\!\lt(\frac{1}{3}\rt)}\frac{1}{\tau}.
\eeq

\subsubsection{Reconstruction of the full solution}
\label{sec:full_sln}

Now let us describe the full solution for the Fourier--Hermite spectrum that 
follows from what we have just derived. 
The spectrum at $\xx\to0$ ($k\to 0$) is given by \exref{eq:Y}. 
In our original variables, this means 
\beq
F_{k\to0}(\tau) = \frac{1}{\tau^{2/3}}
\quad\Rightarrow\quad
C_{k\to0,m} = \frac{\const}{m^{3/2}}. 
\label{eq:Cm}
\eeq
The spectrum at $\tau\to0$ (low $m$) is, via \exref{eq:Fm0} and \exref{eq:Y}
(changing integration variable to $z=\xx^2/4\sigma$), 
\beq
F_0^\pm(\xx) = \frac{2^{4/3}}{\Gamma\!\lt(\frac{1}{3}\rt)}\frac{1}{\xx^{4/3}} 
\quad\Rightarrow\quad
F_k(\tau\to0) = \frac{3^{4/3}}{\Gamma\!\lt(\frac{1}{3}\rt)}\frac{\gamma^{2/3}}{k^2}
\quad\Rightarrow\quad
C_{k,m\to\mmin} = \frac{\const}{k^2}.  
\label{eq:Ck}
\eeq
These two scaling laws, the Hermite spectrum \exref{eq:Cm} and the 
Fourier spectrum \exref{eq:Ck} also hold asymptotically across 
the entire phase space, at large enough $m$ and $k$, as we shall see shortly. 

\begin{figure}
\centerline{\includegraphics[width=0.8\textwidth]{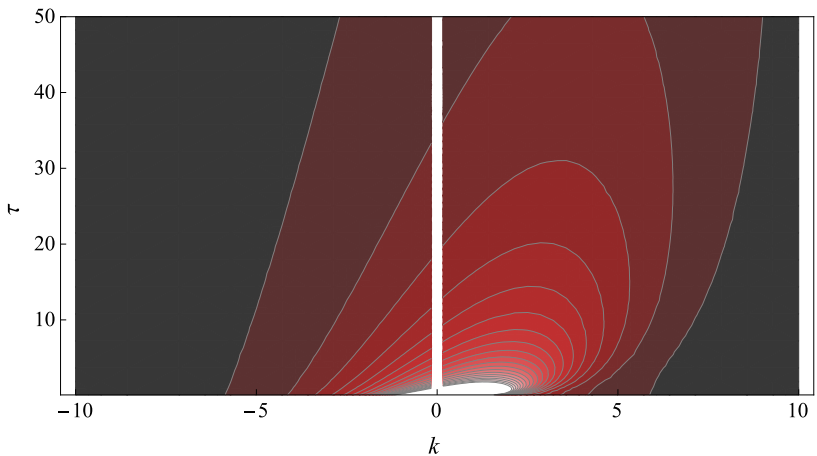}} 
\caption{The function $F_k(\tau)$ given by \exref{eq:Fkm_final} (with $\gamma=1$). 
The phase-mixing 
part of the spectrum is on the right ($k>0$), the anti-phase-mixing part on the 
left ($k<0$). The total energy is the sum and the Hermite flux the difference 
of these two. The broader-range scaling behaviour is better represented 
on a log scale: see \figsref{fig:Fkm_cuts} and \ref{fig:Gkm}.}
\label{fig:Fkm}
\end{figure}

Using \exref{eq:Y} in \exref{eq:Fm_sln} and changing the integration 
variable to $z=\xx^2/4(\sigma-\tau)$, we find 
\beq
F^-(\tau,\xx) = \frac{1}{\Gamma\!\lt(\frac{1}{3}\rt)}
\int_0^\infty\rmd z\,\frac{e^{-z}}{\lt(\tau z + \frac{1}{4}\xx^2\rt)^{2/3}}
= \frac{e^{\xx^2/4\tau}}{\tau^{2/3}}\frac{\Gamma\!\lt(\frac{1}{3};\frac{\xx^2}{4\tau}\rt)}{\Gamma\!\lt(\frac{1}{3}\rt)}.
\label{eq:Fm_final}
\eeq
In the first of these expressions, both asymptotics \exref{eq:Cm} and \exref{eq:Ck} 
are manifest. In the second expression, which is obtained by changing the 
integration variable $z + \xx^2/4\tau\to z$, 
\beq
\Gamma\!\lt(\frac{1}{3};\frac{\xx^2}{4\tau}\rt) = 
\int_{\xx^2/4\tau}^\infty\rmd z\,z^{-2/3} e^{-z}
\eeq 
is an upper incomplete gamma function. 

Finally, using \exref{eq:Ck} and \exref{eq:Y} in \exref{eq:Fp_sln}, we find
\begin{align}
\nonumber
F^+(\tau,\xx) &=\frac{e^{-\xx^2/4\tau}}{\tau^{2/3}}\lt\{1 + (-1)^{2/3}\Biggl[
\frac{\Gamma\!\lt(\frac{1}{3};-\frac{\xx^2}{4\tau}\rt)}{\Gamma\!\lt(\frac{1}{3}\rt)}
-1\Biggr]\rt\}\\
&= \frac{e^{-\xx^2/4\tau}}{\tau^{2/3}}\lt[1 + 
\frac{1}{\Gamma\!\lt(\frac{1}{3}\rt)}
\int_0^{\xx^2/4\tau}\rmd z\,z^{-2/3} e^z\rt].
\label{eq:Fp_final}
\end{align}
The first term inside the bracket comes from the second integral in \exref{eq:Fp_sln}
(the boundary-value term) and is obtained by the manipulations analogous to those that 
led to \exref{eq:Fm_final}. The second term comes from the first integral 
in \exref{eq:Fp_sln} (the initial-value term), which is turned into a tabulated 
integral by changing the integration variable to $z=\xx\yy/2\tau$ (and then 
changing $z\to-z$ to obtain the last integral representation, which is 
perhaps more transparent than the one in terms of the incomplete gamma function). 

\subsubsection{Self-similar solution}
\label{sec:ssim}

\begin{figure}
\begin{center}
\begin{tabular}{cc}
\includegraphics[width=0.475\textwidth]{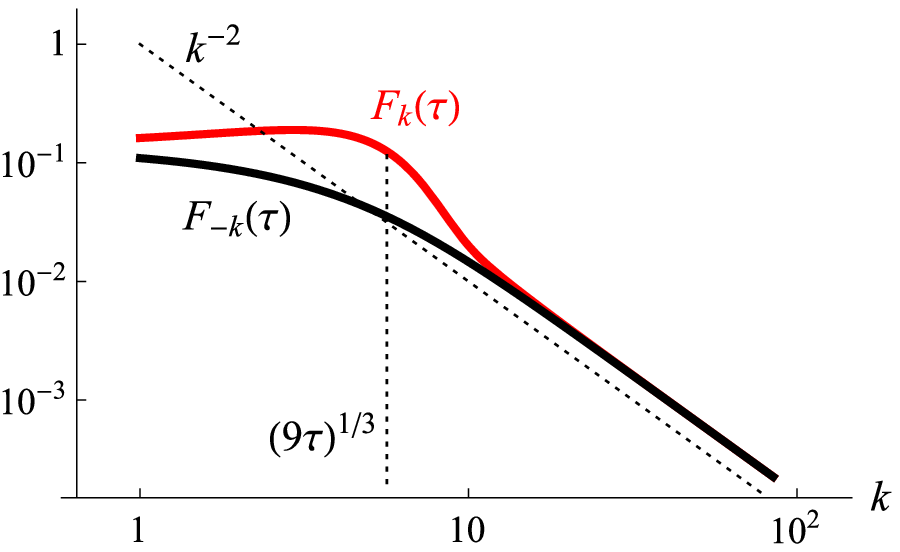}&
\includegraphics[width=0.475\textwidth]{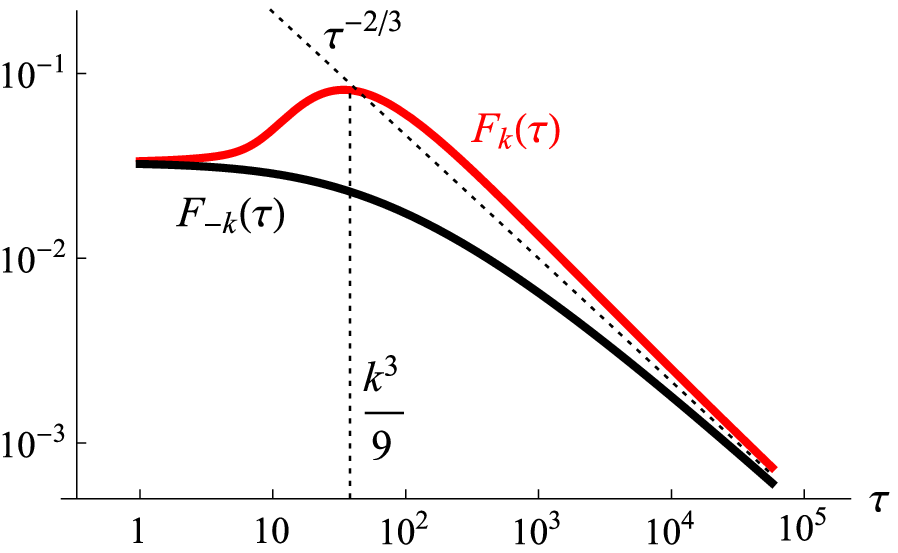}\\\\
(a) $\tau=20$ & (b) $k=7$
\end{tabular}
\end{center}
\caption{Typical cuts through the solution \exref{eq:Fkm_final} (with $\gamma=1$) 
at (a) constant $\tau$, (b) constant $k$. 
The spectra of phase-mixing modes, $F_{k}(\tau)$, are shown in red and 
the spectra of anti-phase-mixing modes, $F_{-k}(\tau)$, in black. 
The asymptotic scalings \exref{eq:Ck} and \exref{eq:Cm} are 
shown for reference and convergence to them is manifest in the limits 
$k\gg(9\tau)^{1/3}$ and $\tau\gg k^3/9$, respectively. 2D plots of $F_k(\tau)$ 
and of the resulting Hermite flux and total energy spectrum are in \figref{fig:Gkm}.}
\label{fig:Fkm_cuts}
\end{figure}

Assembling \exref{eq:Fm_final} and \exref{eq:Fp_final} together and returning 
them to the original variables, we arrive at the following solution:  
\beq
F_k(\tau) = \frac{e^{-k^3/9\gamma\tau}}{\tau^{2/3}}
\lt\{
\begin{array}{l}
\displaystyle
1 + \frac{1}{\Gamma\!\lt(\frac{1}{3}\rt)}
\int_0^{k^3/9\gamma\tau}\rmd z\,z^{-2/3} e^z,\quad k>0,\\\\
\displaystyle
\frac{1}{\Gamma\!\lt(\frac{1}{3}\rt)}
\int_{|k|^3/9\gamma\tau}^\infty\rmd z\,z^{-2/3} e^{-z}, \quad k<0,
\end{array}
\rt.
\label{eq:Fkm_final}
\eeq 
where $\tau = (2m)^{3/2}/3 - \tmin$ ($\tmin$ is an order-unity offset). 
Hence the Fourier--Hermite spectrum $C_{k,m}$ and the Hermite flux $\Gamma_{k,m}$
can be calculated according to \exref{eq:C_F} and \exref{eq:Gamma_F}, 
respectively. 

\begin{figure}
\begin{center}
\begin{tabular}{cc}
\includegraphics[width=0.475\textwidth]{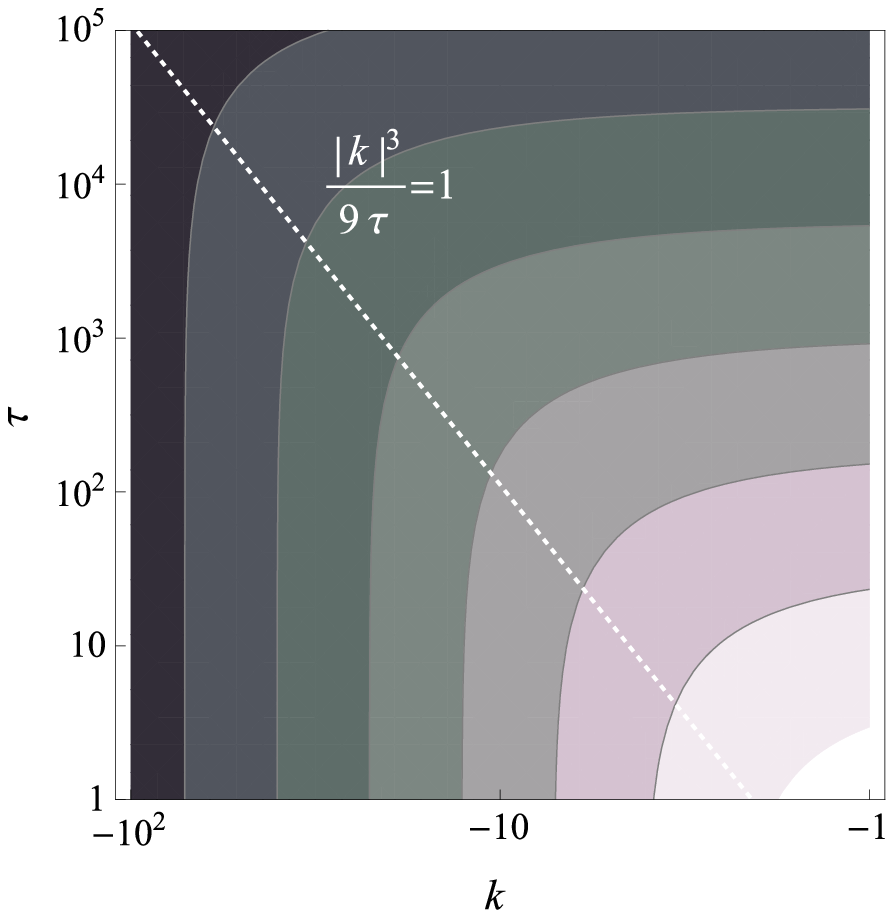}&
\includegraphics[width=0.475\textwidth]{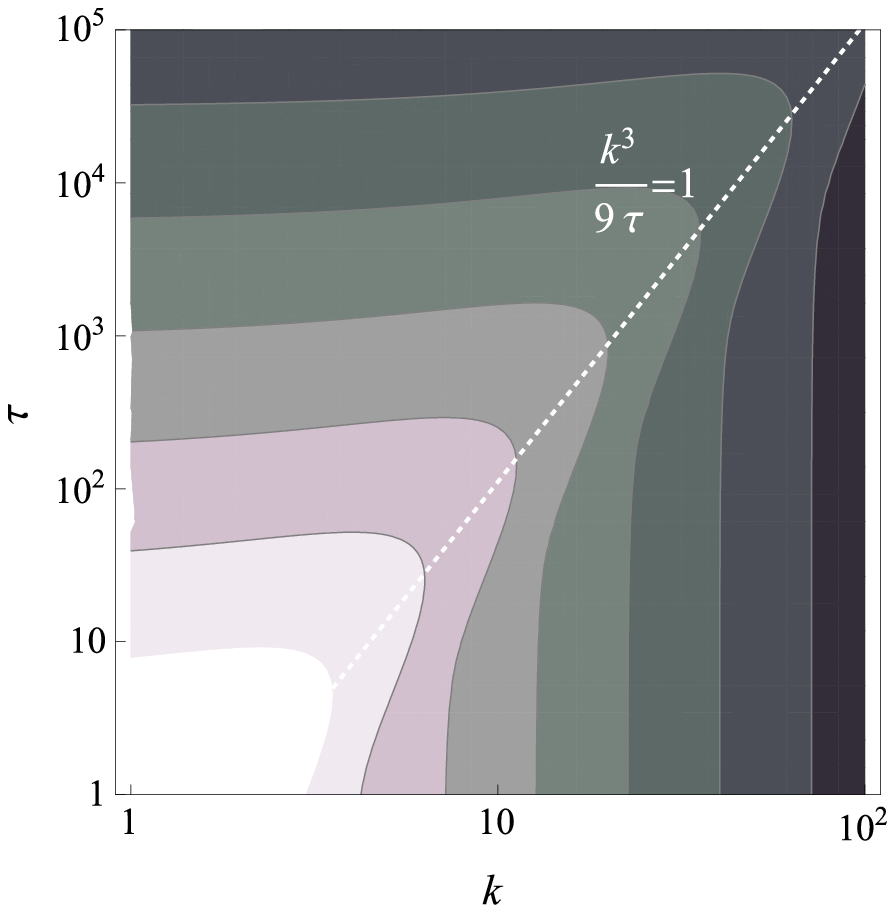}\\
(a) $F_k(\tau)$, $k<0$ & (b) $F_k(\tau)$, $k>0$\\\\\\
\includegraphics[width=0.475\textwidth]{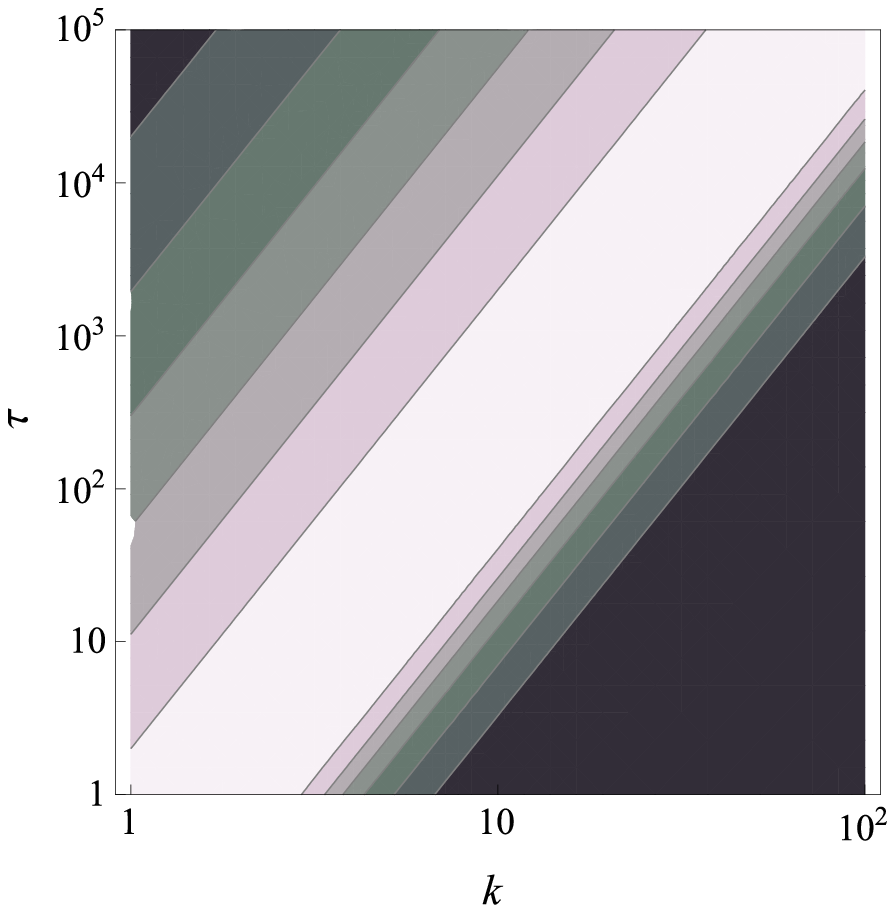}&
\includegraphics[width=0.475\textwidth]{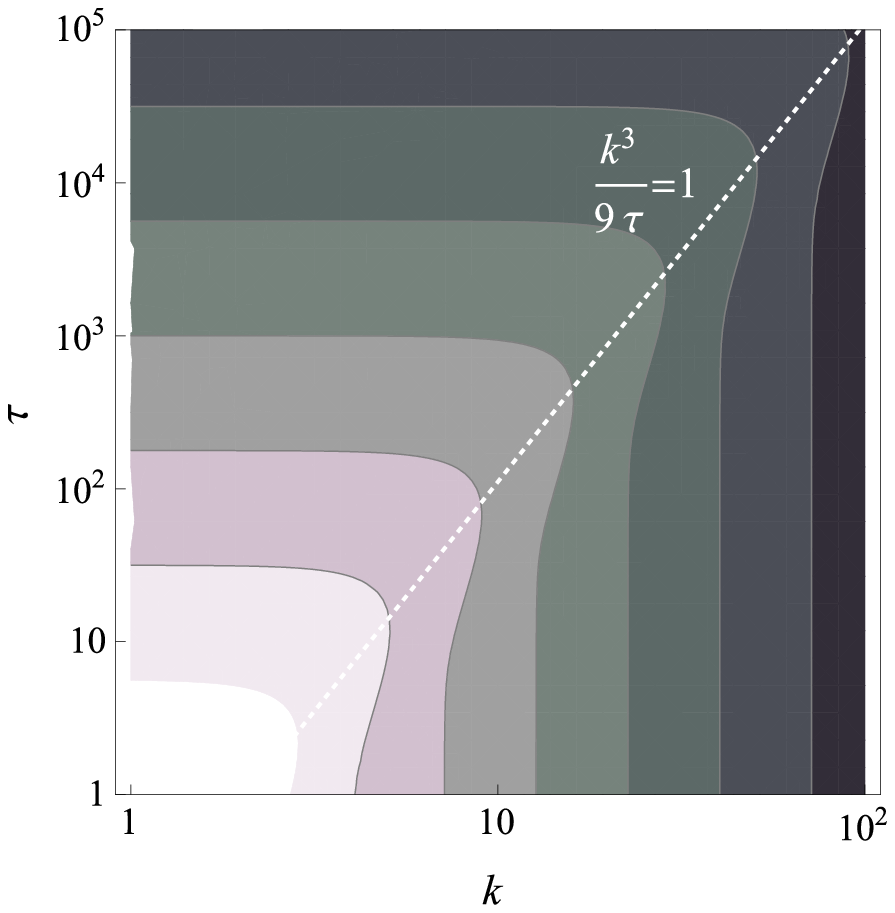}\\
(c) $\bar\Gamma_k(\tau)$ & (d) $\lt[F_k(\tau)+F_{-k}(\tau)\rt]/2$
\end{tabular}
\end{center}
\caption{(a,b) Same as \figref{fig:Fkm}, but on a log scale and across 
a broader range of $k$ and $\tau$. Note the asymptotic features 
(zero-flux solution, power-law scalings) discussed 
in \secsand{sec:ssim}{sec:colls}. Cf.~\figref{fig:cartoon}.
(c) Normalised Hermite flux $\bar\Gamma$ defined by \exref{eq:Gnorm}. 
Its contours are straight in the logarithmic coordinates because 
$\bar\Gamma = \bar\Gamma(k^3/9\tau)$. (d) Total energy spectrum 
$\lt[F_k(\tau)+F_{-k}(\tau)\rt]/2$ [see \exref{eq:C_F}].}
\label{fig:Gkm}
\end{figure}

The solution \exref{eq:Fkm_final} is plotted in \figref{fig:Fkm}, which shows 
a pleasingly nontrivial shape. The essential result is, however, extremely simple. 
The solution is self-similar and could, in fact, have been 
obtained as such, by a shorter, if marginally less general, 
route (see \apref{ap:ssim}). 
The similarity variable $k^3/9\gamma\tau$ determines the demarcation of the 
phase space into two asymptotic regions:  
the asymptotic of the spectrum when $|k|\gg(9\gamma\tau)^{1/3}$ is 
\exref{eq:Ck} and the asymptotic when $\tau\gg|k|^3/9\gamma$ is \exref{eq:Cm}
[this is particularly obvious in the second expression in \exref{eq:Fm_sln}]. 
The former describes fluctuations in the ``wave-number inertial range'' 
with a vanishing Hermite flux, 
the latter fluctuations in the ``Hermite inertial range'', which also have zero 
Hermite flux. These scalings are illustrated in \figref{fig:Fkm_cuts} 
and the normalised Hermite flux
\beq
\bar\Gamma = \frac{F_k(\tau) - F_{-k}(\tau)}{F_k(\tau) + F_{-k}(\tau)} 
= \frac{\Gamma_{k,m}}{\sqrt{2m}\,k C_{k,m}}
\label{eq:Gnorm}
\eeq
is plotted in \figref{fig:Gkm}(c). $\bar\Gamma$ is a good measure of how different 
the nonlinear state is from the linear one: for linear Landau-damped perturbations, 
we would have had $\bar\Gamma = 1$ everywhere \citep{kanekar15}. 
Note that, as follows immediately from \exref{eq:Fkm_final},  
$\bar\Gamma = \bar\Gamma(k^3/9\gamma\tau)$ is a function 
of the similarity variable only.

Another useful result is the overall Hermite spectrum integrated over all wave numbers. 
While this, of course, misses the relationship between structure in position 
and velocity space that we have focused on so closely, it is a good crude measure 
of how ``phase-mixed'' the distribution is \citep[cf.][]{hatch14,servidio17}. 
So, from \exref{eq:C_F} and \exref{eq:Fkm_final}, 
after integrating out the self-similar functional dependence of $F_k(\tau)$ 
on $k^3/9\gamma\tau$, we deduce 
\beq
\sum_k C_{k,m} = \sum_k\frac{F_k + F_{-k}}{2\sqrt{m}} \propto \frac{1}{m}. 
\label{eq:Cm_overall}
\eeq

This scaling---or, equivalently, the $k$-by-$k$ $C_{k,m}\propto m^{-3/2}$ scaling 
at large $m$ [see \exref{eq:Cm}],---being steeper than $m^{-1/2}$, 
implies that our solution does indeed decay fast enough in $m$ in order for the collisional 
dissipation to vanish at vanishing collisionality and so treating the collisionless limit as 
nonsingular was justified (see discussion in \secref{sec:budget}).\footnote{Note, however, 
that \exref{eq:Cm_overall} implies that the amount of energy stored in phase 
space is logarithmically divergent: anticipating the collisional estimates 
in \secref{sec:colls}, we get $\sum_m\sum_k C_{k,m} \propto |\ln\nu|$, by integrating 
up to $m\sim\gamma^{1/3}/\nu$. The same result can be obtained from \exref{eq:Ck_overall} 
by integrating up to $k_\nu\sim (\gamma/\nu)^{1/2}$. This is to be contrasted 
with $\sum_m\sum_k C_{k,m} \propto \nu^{-1/3}$ in the linear regime \citep{kanekar15}.} 
Still, restoring finite $\nu$ leads to a kind of ``Kolmogorov scale'' for our kinetic 
turbulence and to a quantitative measure of the applicability, or otherwise, 
of the linear approximation, so we are going to do this in \secref{sec:colls}.  

Finally, we may also calculate the overall wave-number spectrum of the free energy: 
again integrating out the self-similar functional dependence of $F_k(\tau)$, we get 
\beq
\sum_m C_{k,m} \propto \frac{1}{k}. 
\label{eq:Ck_overall}
\eeq
Since the total variance of the perturbed distribution function 
is conserved in the Kraichnan--Kazantsev model (see \secref{sec:budget}), 
the above result can be made made sense of as the classical \citet{batchelor59} scaling 
of a passive scalar advected by a single-scale stochastic field.

\subsection{Phase-space energy flows and role of collisions}
\label{sec:colls}

\begin{figure}
\centerline{\includegraphics[width=0.8\textwidth]{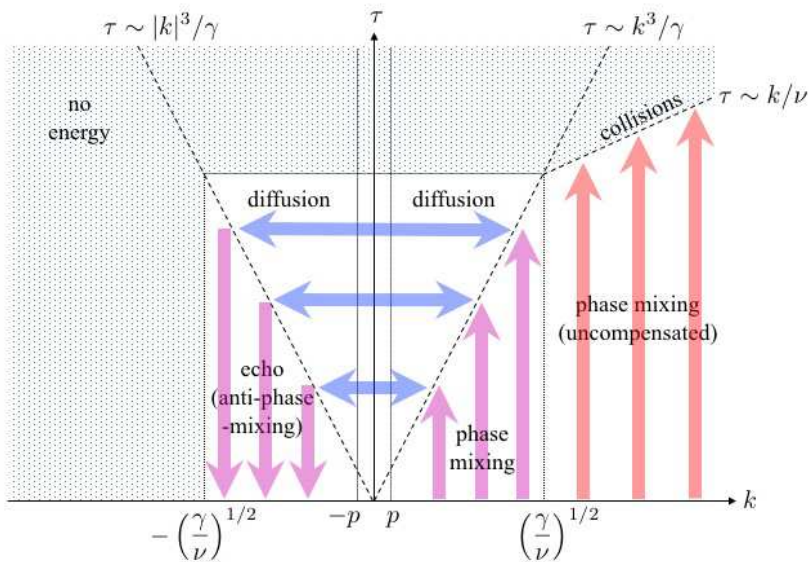}} 
\caption{Cartoon of energy flows in phase space, including collisional cutoff. 
Cf.~\figref{fig:Gkm}(a,b). Reminder: $\tau\sim m^{3/2}$ [see \exref{eq:Fstst}]
and $\gamma$ is related to the stochastic electric field via \exref{eq:tnl}.}
\label{fig:cartoon}
\end{figure}

The structure of our self-similar solution of \exref{eq:Floc} is, in fact, 
easily understood already by means of a qualitative examination of the equation, 
which is also a useful approach in evaluating the role of collisions.  
\Figref{fig:cartoon} is a cartoon of phase-space energy flows in aid of the 
discussion that follows [cf.\ \figref{fig:Gkm}(a,b)]. 

\subsubsection{Phase-mixing region}

The phase-mixing term dominates over the nonlinearity when 
\beq
\tau \ll \frac{k^3}{\gamma} 
\quad\Rightarrow\quad
\frac{\dd F_k}{\dd \tau} = 0,
\label{eq:ph_mix_region}
\eeq
so the solution is independent of $\tau$ in this region [see \figref{fig:Gkm}(b)]. 
This is the linear phase-mixing 
solution, $C_{k,m}\propto 1/\sqrt{m}$, obtained earlier by \citet{zocco11} 
and \citet{kanekar15}. Its wave-number scaling, $F_k\propto 1/k^2$, is, however, 
a new feature, extracting which required matching with other regions. 
  
A way of making sense of this solution is to go back 
to the time-dependent \eqref{eq:tf} [or \exref{eq:Fclosed}] and notice that, for $k>0$, 
whatever solution $\tf_k$ (and, therefore, $F_k$) exists at low $s$, it will 
propagate ``upwards'' (to higher $s$) along the characteristic 
\beq
s = \frac{kt}{\sqrt{2}},
\label{eq:char}
\eeq 
and it will do so unimpeded by the nonlinearity as long as the time $t$ is shorter 
than the nonlinear time $\tnl$ associated with the mode-coupling term in the 
right-hand side of those equations. In the Kraichnan--Batchelor model, 
\beq
\tnl^{-1}\sim \frac{s^2\gamma}{k^2},\quad \gamma \sim p^4\kap_p \sim p^4 \ephi_p^2\tc,
\label{eq:tnl}
\eeq
where $\tc$ is the correlation time of the wave field $\ephi_p$ 
(effectively assumed to be single-scale). 
Requiring $t\ll \tnl$ in \exref{eq:char}, we see that 
the phase-mixing region of the phase space extends to $s^3\ll k^3/\gamma$, 
which is the same as the condition in \exref{eq:ph_mix_region}. 

This gives us a way to estimate how small the wave amplitude must be in order 
for the nonlinearity and associated effects never to matter: indeed, if 
the collision time is short compared to the nonlinear time,
\beq
\tcoll \sim \frac{1}{\nu s^2} \ll \tnl 
\quad\Leftrightarrow\quad
k \gg \lt(\frac{\gamma}{\nu}\rt)^{1/2}\equiv k_\nu,
\label{eq:knu}
\eeq 
the phase-mixed distribution function will thermalise before the echo can 
bring any energy back from phase space. Putting $\tcoll$ into \exref{eq:char} 
tells us how far into phase space energy will travel: 
\beq
\tau \sim s^3 \ll \frac{k}{\nu}\equiv\tau_\nu,\quad k\gg k_\nu. 
\label{eq:snu}
\eeq 

Perhaps the most practically important conclusion from this is that, 
given the strength of the electric field and, therefore, via \exref{eq:tnl}, 
the value of $\gamma$, we can predict the collisional cutoff wave number  
$k_\nu$, given by \exref{eq:knu}, at which phase mixing (Landau damping) 
curtails the universal spectrum that we have derived above: we shall 
see in a moment that for $k\gg k_\nu$, $\bar\Gamma = 1$, i.e., there is 
no echo flux from high to low Hermite moments.   

\subsubsection{Diffusion and echo regions}

Considering the limit opposite to \exref{eq:ph_mix_region}, i.e., the region 
of phase space where $s\gg k\tnl$, we get 
\beq
\tau \gg \frac{k^3}{\gamma} 
\quad\Rightarrow\quad
\frac{\dd^2 F_k}{\dd k^2} = 0, 
\eeq
i.e., $k$-space diffusion dominates. Our self-similar solution (\secref{sec:ssim}) 
tells us that the appropriate 
solution in this region is one independent of $k$ [see \figsref{fig:Gkm}(b) and \ref{fig:Gkm}(a)] 
and $F_k(\tau)\propto 1/\tau^{2/3}$. This solution takes whatever values $F_k(\tau)$ 
has at $k\sim (\gamma\tau)^{1/3}$ [\figref{fig:Gkm}(b)] and transfers them across 
to $k\sim -(\gamma\tau)^{1/3}$ [\figref{fig:Gkm}(a)], where anti-phase-mixing 
picks them up and transfers them ``downwards'' to low Hermite moments ($\tau\to0$), 
over times that are again shorter than $\tnl$ (because $\tau \ll |k^3|/\gamma$ again), 
along the characteristic 
\beq
s = s_0 - \frac{|k|t}{\sqrt{2}},
\eeq
where $s_0\sim \tau_0^{1/3} \sim |k|/\gamma^{1/3}$. 
 
This last piece of the solution 
is the echo flux. It cancels the phase-mixing flux exactly, provided 
the energy ($F_k$) from $k>0$ has been successfully transferred by 
phase mixing from $\tau=0$ to $\tau \sim k^3/\gamma$ to be picked up 
by diffusion and carried over to the anti-phase-mixing region $k<0$. 
For $k\gg k_\nu$, the energy gets intercepted at $\tau\sim k/\nu$ 
[see \exref{eq:snu}] and thermalised by collisions, so at $\tau\sim k^3/\gamma$, 
$F_k(\tau)=0$. This then gives $F_{-k}(0)=0$, i.e., no echo flux. 
Thus, $k_\nu$ is indeed the wave-number cutoff---a kind of ``Kolmogorov scale'' for 
Vlasov-kinetic turbulence---beyond which 
Landau damping can act as an efficient route to (eventually collisional) dissipation. 

\section{Discussion}
\label{sec:disc}

\subsection{Summary}

We have considered what is arguably the simplest kinetic turbulence problem 
available: a 1D Vlasov--Poisson 
(\secref{sec:emodel}; or Vlasov--Boltzmann: see \secref{sec:imodel}) plasma
with an energy source. When collisions are vanishingly weak, this gives rise 
to interesting dynamics across the 2D (position and velocity) phase space. 
In a simple approximation where the stochastic electric field mixing the particle 
distribution can be assumed to have statistics independent of the 
high-order moments of this distribution, a solvable model can be constructed 
in the same vein as the Kraichnan--Batchelor model used in the passive-scalar 
problem. The resulting analytical solution displays the same key features 
as have been surmised heuristically \citep{sch16} and found numerically 
\citep{parker16} for plasma systems with higher-dimensional 
phase spaces (see \secref{sec:3D}). 

Namely, the free-energy flux from low to high Hermite moments 
is suppressed---i.e., the dissipation channel associated with 
Landau damping is shut down---for all 
wave numbers below a certain cut off $k_\nu = (\gamma/\nu)^{1/2}$ 
(see \secref{sec:colls}). This cut off is a kinetic analog of the 
Kolmogorov scale: it scales inversely with the collision rate $\nu$ 
and increases with the amplitude of the electric perturbations---the latter 
determines $\gamma$, which is the rate of diffusion of the free 
energy in $k$ space due to the stochastic electric field. 
Thus, one might expect a kind of statistical ``fluidisation'' 
of the turbulence in the ``inertial range'' ($k\ll k_\nu$)---perhaps 
a welcome development from the point of view of the long history of attempts 
to reduce kinetics to fluid (or ``Landau-fluid'') dynamics 
(see references and further discussion in \secref{sec:LF}). 

Expanding our interest beyond the effect of phase-space turbulence on the 
low (``fluid'') moments of the distribution function and to the structure 
of this turbulence across the (Fourier--Hermite) phase space, 
we find the latter cleanly partitioned into two regions: 
(i) the {\em phase-mixing region} $k\gtrsim \gamma^{1/3}\sqrt{m}$, 
where phase mixing and anti-phase-mixing transfer free energy between 
lower and higher Hermite moments (cancelling on average), 
and (ii) the {\em mode-coupling (or diffusion) region} $k\lesssim \gamma^{1/3}\sqrt{m}$, 
where the free energy is transferred between spatial scales 
(wave numbers) by the advecting action of the stochastic electric field. 
An overview of how this happens is provided in \secref{sec:colls} 
and \figref{fig:cartoon}, while the Fourier--Hermite spectrum is 
derived more formally in \secref{sec:solution} (summarised in 
\secref{sec:ssim}). The resulting scalings are 
\beq
C_{m,k} \sim \lt\{
\begin{array}{ll}
\displaystyle
\frac{1}{\gamma^{2/3} m^{3/2}}, & 
\displaystyle
k \lesssim \gamma^{1/3}\sqrt{m},\quad m \ll \frac{\gamma^{1/3}}{\nu},\\\\
\displaystyle
\frac{1}{k^2\sqrt{m}}, & 
\displaystyle
\gamma^{1/3}\sqrt{m}\lesssim k \ll k_\nu = \lt(\frac{\gamma}{\nu}\rt)^{1/2}.
\end{array}
\rt.
\label{eq:Cmk_summary}
\eeq

\subsection{Open issues} 

There is a number of questions and lines of investigation that all this leaves open. 
The more immediate and obvious of them are, naturally, to do with how 
universal these results are, given the radical nature of the approximations  
that were made in order to obtain them, and how they can be made more general 
and more applicable to concrete physical problems. 
Let us itemise these questions briefly. 

\subsubsection{Multiscale electric fields}

What happens outside Batchelor's approximation (\secref{sec:Batchelor}), 
i.e., when the stochastic electric field cannot be treated as effectively single-scale 
and, therefore, as providing a diffusively accumulating series of small kicks 
in $k$ space to the perturbed distribution function? Formally, dealing with 
this issue is a matter of solving the integral equation \exref{eq:Fstst}, 
rather than the differential equation \exref{eq:Floc}. A certain (limited) 
amount of progress on this 
is made in \apref{ap:Zslns}, suggesting somewhat steeper wave-number spectra 
in the mode-coupling region. While we do not have the full solution, 
it appears plausible that, even though scalings might change, 
the overall partitioning of the phase space into (anti-)phase-mixing- 
and mode-coupling-dominated regions should persist and the Landau damping 
would still be suppressed. In fact, the conversion between phase-mixing 
and anti-phase-mixing modes may be quicker in this case than in Batchelor's 
limit because the steps in $k$ space need not be as small as  
in the diffusive regime \citep[cf.][]{sch16}. 

\subsubsection{Finite-time-correlated electric fields}

What happens outside Krachnan's approximation (\secref{sec:KK}), 
i.e., if the assumption of a short correlation time of the electric field is 
relaxed? The experience of passive-advection problems in fluid dynamics 
\citep[see, e.g.,][]{antonsen96,bhat15} suggests that generalising 
from white to finite-time-correlated advecting field 
rarely leads to dramatic qualitative changes. In the context of plasma 
kinetics, however, the limit, opposite to ours, of a long correlation time 
of the electric field clearly requires special care to capture the phase-space 
structure that will arise due to particle-trapping effects 
\citep{bernstein57,oneil65,oneil71,manheimer71}.   

\subsubsection{Self-consistent electric fields}

How does one construct a fully nonlinear theory of kinetic turbulence, i.e., 
one in which the stochastic electric field is not prescribed but is 
determined self-consistently? This requires coupling the high-$m$ 
dynamics to the low-$m$ ``fluid'' equations derived in \secref{sec:fluid}, 
i.e., one would have to progress from the vague understanding that the 
flux to/from high Hermite moments is suppressed to a more quantitative 
model of how this suppression can be incorporated into a dynamical 
(``Landau fluid''; \secref{sec:LF}) or, more likely, statistical model 
of the low-$m$ moments (and how many of them must be kept). Once this is done, 
it becomes possible to assess to what extent a self-consistent electric field 
can be compatible with the modelling choices made above: the Batchelor (single scale)
and Kraichnan (short correlation time) approximations. 

Without claiming to have such a theory, let us offer 
a few na\"ive but plausible estimates. Taking the limit of our solution 
at low $m$ [the second asymptotic in \exref{eq:Cmk_summary}], we 
find a $k^{-2}$ spectrum. Let us conjecture that this spectrum 
is established in phase space and imposes itself, via the linear Hermite 
coupling term in \exref{eq:gm}, on the lower Hermite moments.\footnote{It is 
as yet poorly understood how the separation between the ``fluid'' Hermite 
moments and the ``kinetic'' ones should be made, i.e., how many lower moments 
ought to be treated as discrete, distinct fluid-like fields and starting 
with what $m$ the formalism based on continuity in Hermite space 
(\secref{sec:high_m} and onwards) can safely take over. It seems clear that 
this transition must be at $m$ of order unity (and certainly independent 
of $\nu$), except, in view of \exref{eq:cond_cont}, when $\alpha_k\gg1$ 
(i.e., when the wave frequency $\omega_k$ is large compared to the particle 
streaming rate $k\vth$). In the latter case, we cannot use the $1/\sqrt{m}$ 
Hermite spectrum until $m\sim\alpha_k/4$ (in the linear theory, a different 
spectrum can be derived for $1\ll m \ll \alpha_k/4$, which falls off 
in great leaps from one $m$ to the next until it morphs into $1/\sqrt{m}$
at $m\sim\alpha_k$; see \citealt{kanekar15}, \S4.3).} 
So, starting with \exref{eq:gm} for $m=3$, we may 
predict\footnote{Note that an extreme way to achieve the balancing of 
the phase-mixing and anti-phase-mixing fluxes is to suppress 
every other Hermite moment, e.g., the odd ones (starting with 
the heat flux $g_{k,3}$, this would indeed shut down Landau damping completely).}
\beq
\la|T_k|^2\ra \equiv 2\la|g_{k,2}|^2\ra \sim \la|g_{k,4}|^2\ra 
\sim \la|g_{k,6}|^2\ra \sim \dots \sim k^{-2}. 
\label{eq:Tk_spectrum}
\eeq 
From the wave equation \exref{eq:npm} and in view of \exref{eq:npm_def}, we may estimate 
\beq
n_k^\pm \sim \frac{k^2}{\omega_k^2}\,\theta_k = \frac{2}{3+\alpha_k}\,\theta_k,
\quad  
n_k \sim n_k^\pm,\quad
\uu_k \sim \frac{\omega_k}{k}\, n_k^\pm = \sqrt{\frac{3+\alpha_k}{2}}\,n_k^\pm.
\eeq
Note that, regardless of the size of $\alpha_k$, $n_k$ is always either much 
smaller than or comparable to $\theta_k$. Therefore, $T_k = \theta_k + 2n_k$ 
in \exref{eq:Tk_spectrum} can be replaced by $\theta_k$ for the purposes 
of these crude estimates. This gives us a prediction for the spectra 
of the ``wave quantities'': 
\beq
\la|n_k|^2\ra \sim \lt(\frac{2}{3+\alpha_k}\rt)^2 k^{-2},\quad
\la|\uu_k|^2\ra \sim \frac{2}{3+\alpha_k}\, k^{-2}.
\eeq 
At wave numbers where, in \exref{eq:phik}, the self-consistent term 
($\alpha_k\ephi_k$) dominates over the ``external'' one ($\chi_k$), 
the electric-field spectrum~is 
\beq
\la|E_k|^2\ra = k^2\la|\ephi_k|^2\ra \sim k^2 \alpha_k^2 \la|n_k|^2\ra
\sim \lt(\frac{2\alpha_k}{3+\alpha_k}\rt)^2. 
\eeq
This tells us that, for example, for Vlasov--Poisson perturbations 
with $k\lDe\gg1$ and $\alpha_k= 2/k^2$ (see \secref{sec:emodel}), 
we should have $\la|E_k|^2\ra\sim k^{-4}$, 
which is a steep enough scaling to justify Batchelor's approximation.\footnote{Only 
just steep enough: we find $K_k = k^2\kap_k \sim \la|E_k|^2\ra\tc \sim k^{-2}$ if we estimate 
the correlation time, from \exref{eq:npm}, 
as $\tc^{-1}\sim (k/\omega_k)k\ephi_k \sim \la|E_k|^2\ra^{1/2}\sqrt{2/(3+\alpha_k)}$.} 
In contrast, if $\alpha_k$ is 
a constant (ion-acoustic perturbations; see \secsand{sec:imodel}{sec:Zmodel}), 
the self-consistent $\la|E_k|^2\ra$ is flat (cut off at $k\sim k_\nu$)
and our model can only work if the external potential $\chi_k$ 
plays the dominant part in the nonlinear mode coupling 
(in the case of ion-scale Zakharov turbulence, \secref{sec:Zmodel}, 
this might be a credible possibility as $\chi_k$ is fed by the 
fast-oscillating electric fields that satisfy an electron-time-scale 
fluid-like equation and have a tendency to form a large-scale condensate; 
see \citealt{zakharov72} and the reviews cited in \secref{sec:Zmodel}). 

Finally, if one is interested in the stochastic-acceleration problem 
($\alpha_k=0$; \secref{sec:stoch_acc}), the electric field need not 
be self-consistent and the conclusion from the above considerations 
is that particles stochastically accelerated by a short-time-correlated
electric field cut off above wave number $p$ will develop $k^{-2}$ 
density, velocity, temperature, etc.\  spectra at $k\gg p$. 
This appears to be a new result. 

The above discussion should probably not satisfy a discerning reader: 
clearly, there is much to be done before the link between 
the ``kinetic'' and ``fluid'' moments is established in an analytically 
solid and quantitative way. 

\subsubsection{Implications for Landau-fluid closures} 
\label{sec:LF}

The question of what constitutes a quantitatively accurate closure scheme for the first 
few fluid moments of a kinetic plasma system has been studied in some detail, 
both analytically and numerically, in the framework of ``Landau-fluid'' closures 
\citep{hammett90,hammett92,hammett93,dorland93,beer96,smith97,snyder97,passot04,goswami05,tassi16,passot17} 
and, indeed, by their detractors \citep[e.g.,][]{mattor92,weiland92}.
A crude view of the philosophy behind this approach is that linear 
Landau-damping rates are incorporated explicitly into low-$m$ fluid equations 
to model the energy removal process into higher-order moments. This might 
appear to be inconsistent with the notion, advocated by us, 
that phase mixing is cancelled 
by anti-phase-mixing and thus Landau damping is effectively suppressed 
as a route to thermalisation of the low-$m$ energy. However, the situation 
is, in fact, more nuanced. 

It was understood already in the course of 
development of early Landau-fluid models that their performance improves 
if more fluid moments are included. It was also understood that the cause 
of this improvement is that nonlinearities in those fluid equations 
act to reduce the phase-mixing rate to higher-order ``unresolved'' 
moments \citep{beer96,smith97}. It seems plausible that keeping the ``right'' number 
of moments is tantamount to keeping enough $m$'s to enable the system to have 
enough anti-phase-mixing to capture the cancellation effect reasonably well---and 
that if the truncation is done at some $m$ that is still smaller than the 
$m$ at which the flux into higher-order moments is fully cancelled (which 
will depend on $k$), 
some form of Landau-fluid closure may be adequate to mop up the residual flux. 

Furthermore, if the cut-ff scale $k_\nu$ is finite, i.e., if 
the collision rate is sufficiently large (equivalently, the velocity-space 
resolution of a code is limited) and/or the fluctuation amplitude 
is sufficiently small, the dissipation at $k\gtrsim k_\nu$ should be 
perfectly well described by the Landau rate corresponding to those 
$k$ (see \secref{sec:colls}). 
We also saw (\figref{fig:Gkm}c) that, at least in our model, 
cancellation of the Hermite flux required $k\gg p$, i.e., 
worked in the ``inertial range'' ($p\ll k\ll k_\nu$), rather than 
at the ``energy-containing scale'' of the turbulence. It is easy to 
imagine situations (e.g., in near-marginal tokamak turbulence) in which 
the scale separation between $p$ and $k_\nu$ might not be large. 

Thus, the usefulness of Landau-fluid closures, judiciously applied, 
is not obviated by the stochastic-echo effect---but it is clearly an interesting 
topic for exploration how the type of results reported above might 
help one hone these closures with this effect explicitly in one's sights. 

\subsubsection{3D}
\label{sec:3D}

What happens when the system is allowed to be 3D in both velocity and position space? 

In magnetised, {\em drift-kinetic} plasmas, this, in fact, brings 
in only one extra phase-space variable ($\kperp$ in addition to 
$\kpar$, all statistics being isotropic in the plane perpendicular to the magnetic field);  
phase mixing in drift kinetics is only in $\vpar$, 
so there is only one velocity-space dimension---provided the equilibrium 
is isotropic (if it is not, there is also phase-mixing in $\vperp$; 
see \citealt{dorland93}, \citealt{mandell17}). 
The nonlinearity in a magnetised plasma is of a more traditional ``fluid'' 
type, viz., it is the advection of $\df$ by turbulent $\vE\times\vB$ flows.
A certain amount of progress has been made \citep{hatch14,kanekar15phd,sch16,parker16}
and is being made with this problem in application both 
to laboratory-inspired turbulence models and to the solar wind. 

At scales where the finite size of particles' Larmor orbits is felt, 
this triggers vigorous nonlinear phase mixing in $\vperp$: the ``entropy cascade'' 
introduced by \citet{sch08,sch09} (but anticipated already by \citealt{dorland93}) 
and first numerically diagnosed by \citet{tatsuno09} and \citet{banon11prl}
(see also \citealt{kawamori13}, who claims experimental confirmation). 
Thus, {\em gyrokinetic} turbulence has a 5D phase space. Outside the gyrokinetic 
approximation, at high frequencies (comparable and exceeding 
the Larmor frequencies of the particles), the phase space finally becomes 6D, 
as the distribution function can develop structure also in the gyroangle. 

In a recent promising development,  
the first harbinger has appeared \citep{servidio17} of a stage 
in spacecraft exploration of plasma turbulence in the Earth's magnetosphere and the 
solar wind when 3D Hermite spectra (and, perhaps, 6D Hermite--Fourier ones) 
become measurable and thus 
so much more attractive as a subject for theoretical prediction, 
now falsifiable (and not necessarily just Hermite spectra; see \citealt{howes17} 
and \citealt{klein17}).\\ 

This is a good note to finish on: phase-space turbulence as a new frontier 
for observation and measurement, as well as theory---for we must hope that 
the phenomena that are revealed by simple solvable models of kinetic 
turbulence are not just interesting or aesthetically pleasing but also real. 

\begin{acknowledgments} 
We are grateful to T.~Antonsen, F.~Califano, P.~Dellar, 
R.~Meyrand, J.~Parker, D.~Ryutov, J.~Squire and L.~Stipani 
for discussions of this and related problems, and especially to 
W.~Dorland and G.~Hammett, who offered detailed comments on the manuscript. 
T.A.'s work was supported by the R.~Peierls Centre's and Merton College's undergraduate 
summer research bursary schemes. A.A.S.'s work was supported in part 
by grants from UK STFC and EPSRC. 
Both authors gratefully acknowledge the hospitality of the Wolfgang Pauli Institute, 
University of Vienna, where a significant part of this research was performed. 
\end{acknowledgments} 

\begin{appendix}

\section{Solutions of \exref{eq:Fstst} and the Batchelor approximation}
\label{ap:Zslns}

Here we explore how the solutions of \exref{eq:Fstst} might change if the electric-field 
correlation function $K_p= p^2\kap_p$ does not decay quickly enough in order for 
the Batchelor approximation [i.e., the expansion \exref{eq:Batchelor_limit}] to be 
legitimate---and also under what condition on $K_p$ it {\em is} legitimate. 
Even if it cannot be turned into a differential operator,  
the mode-coupling integral in the right-hand side 
of \exref{eq:Fstst} will, qualitatively, still provide a kind of smoothing effect in $k$ space 
(in the Batchelor limit, this was diffusion) and transfer $F_k$ from 
positive to negative wave numbers, causing anti-phase-mixing. One might again 
expect that this mode-coupling term will dominate over the phase-mixing 
term, $k \dd F_k/\dd\tau$, at sufficiently small $k$ and sufficiently large $\tau$, 
whereas in the opposite limit, the phase mixing (or anti-phase-mixing) 
will simply transfer energy ``vertically'' in the $(k,\tau)$ plane 
(analogously to \figref{fig:cartoon}). The solutions of \exref{eq:Fstst} 
in the mode-coupling-dominated region will then be such functions $F_k$ 
that the right-hand side of \exref{eq:Fstst} vanishes. 

It is convenient to write this as follows, converting the wave-number sum into an integral  
\beq
\int_{-\infty}^{+\infty}\rmd q\, K_{k-q}(F_k - F_q) = 0. 
\label{eq:intKF}
\eeq
Let us assume that $K_p\propto |p|^\alpha$ and $F_k\propto |k|^\beta$. 
Scaling out $|k|^{\alpha+\beta+1}$ and changing the integration variable to $\xi=q/|k|$, 
we find that the above integral is proportional to 
\beq
I = \int_0^\infty\rmd \xi\,(1-\xi^\beta)\lt(|1-\xi|^\alpha + |1+\xi|^\alpha\rt). 
\eeq
The change of variables $\eta=1/\xi$ (a 1D Zakharov transformation; see \citealt{zakharov92})
turns this integral into 
\beq
I = -\int_0^\infty\frac{\rmd \eta}{\eta^{2+\beta+\alpha}}(1-\eta^\beta)\lt(|1-\eta|^\alpha + |1+\eta|^\alpha\rt). 
\eeq 
This is just $-I$ and so $I=-I=0$, as required,~if
\beq
\beta = -2 -\alpha
\quad\Rightarrow\quad 
F_k\propto |k|^{-2-\alpha}.
\label{eq:Zslns}
\eeq
This is the desired solution satisfying \exref{eq:intKF}. 
It is only valid provided the integral $I$ converges, the conditions for which are 
\beq
\beta > -1, \quad -2 < \alpha < -1 
\quad\Rightarrow\quad 
-1 < \beta < 0. 
\eeq
Otherwise, the integral \exref{eq:intKF} 
is dominated by what happens at the low- or high-wave-number 
cutoffs or at $|p|=|k-q|\ll|k|$ (i.e., $\xi\approx1$). 

The latter wave-number range ($|p|\ll |k|$) 
takes over the integral \exref{eq:intKF} for $\alpha\le-2$. 
This corresponds to the flat solution ($\beta=0$) at $|k|\lesssim |p|$ that was derived 
in \secref{sec:bc} 
and that at $|k|\gg |p|$ transitions seamlessly into the Batchelor-limit 
solution worked out in \secref{sec:solution} (or \apref{ap:ssim}). 
Thus, the condition of validity of the Batchelor approximation 
is that $K_p$ must certainly decay more steeply than $|p|^{-2}$. 

Whether the shallow power-law solutions \exref{eq:Zslns} have any interesting 
physical applications remains to be seen. Presumably, they indicate 
that when the Batchelor approximation is broken, the $k$ spectrum 
in the mode-coupling-dominated region becomes a little steeper than 
the flat Batchelor-limit solution---with some attendant change in the 
$\tau$ spectrum (and also in the $k$ spectrum in the phase-mixing-dominated 
region), to find which we would need to calculate the solution 
of \exref{eq:Fstst} more precisely, taking into account the fact 
the power law \exref{eq:Zslns} does not, in fact, extend to 
arbitrarily large $k$. 

It is perhaps worth noting that these solutions 
must be rather particular to the 1D world, as wave-number integrals 
expressing nonlinear mode coupling become quite different 
in more dimensions \citep[see, e.g.,][]{zakharov92}. 

\section{Self-similar solution of \exref{eq:Floc}} 
\label{ap:ssim}

Consider \exref{eq:Floc} with $\nu=0$: 
\beq
k\frac{\dd F_k}{\dd\tau} = \gamma\frac{\dd^2 F_k}{\dd k^2}. 
\label{eq:Fk_bare}
\eeq 
This equation has a self-similar solution of the form 
\beq
F_k(\tau) = \frac{1}{\tau^\al}\,\Phi(\kk),\quad \kk = \frac{k^3}{\gamma\tau},  
\label{eq:Fk_ssim}
\eeq
where $\Phi(\kk)$ satisfies an ordinary differential equation readily obtained 
by substituting \exref{eq:Fk_ssim} into \exref{eq:Fk_bare}: 
\beq
9\kk \Phi'' + (6+\kk) \Phi' + \al\Phi = 0. 
\label{eq:Phi}
\eeq

A standard method for fixing the exponent $\al$ for self-similar solutions 
such as \exref{eq:Fk_ssim} is to use some conservation law that the solution 
must satisfy. In this case, there is indeed a conservation law:  
\exref{eq:Fk_bare} implies 
\beq
\frac{\rmd}{\rmd\tau}\int_{-\infty}^{+\infty}\rmd k\,kF_k = 0. 
\label{eq:flux_cons}
\eeq
Using \exref{eq:Fk_ssim}, we find 
\beq
\int_{-\infty}^{+\infty}\rmd k\,kF_k = \frac{\tau^{\frac{2}{3} - \al}}{3}
\int_{-\infty}^{+\infty}\rmd\kk\,\kk^{-1/3}\Phi(\kk), 
\label{eq:int_kFk}
\eeq
which would appear to require 
\beq
\al = \frac{2}{3} 
\eeq
to make the integral \exref{eq:int_kFk} independent of $\tau$. In fact, we know from 
the argument in \secref{sec:budget} that this integral must be zero at $\tau\to0$.   
In view of \exref{eq:flux_cons}, it must then also be zero at all $\tau$. 
Thus, the choice of $\al=2/3$ cannot really be justified {\em a priori}, but it is 
intriguing because with this value, \exref{eq:Phi} is rendered easily solvable, 
so we will examine its consequences before discussing its legitimacy. 

With $\lambda=2/3$, \exref{eq:Phi} is solved by the following substitution:    
\beq
\psi = \Phi' + \frac{1}{9}\Phi
\quad\Rightarrow\quad
\kk \psi' + \frac{2}{3}\psi = 0
\quad\Rightarrow\quad
\psi = \frac{C_1}{|\kk|^{2/3}},   
\eeq
where $C_1$ is an integration constant. Integrating again, we find 
\beq
\Phi(\kk) = e^{-\kk/9}\lt(C_1\int_{-\infty}^\kk \rmd\kk' |\kk'|^{-2/3} e^{\kk'/9} + C_2\rt),
\eeq
where $C_2$ is another integration constant. It is clear that $C_2=0$, lest 
$\Phi(\kk)$ blow up at $\kk\to-\infty$. Cleaning up the remaining solution 
by changing the integration variable to $z=\kk'/9$,
setting $C_1=\lt[9^{1/3}\Gamma(1/3)\rt]^{-1}$ (this is arbitrary and done for aesthetic reasons) 
and using the resulting $\Phi(\kk)$ in \exref{eq:Fk_ssim}, we arrive at the following solution 
\beq
F_k(\tau) = \frac{e^{-k^3/9\gamma\tau}}{\tau^{2/3}}
\frac{1}{\Gamma\!\lt(\frac{1}{3}\rt)}\int_{-\infty}^{k^3/9\gamma\tau} \rmd z\,|z|^{-2/3} e^z.
\label{eq:ssim_sln}
\eeq
If we split the $k>0$ and $k<0$ cases explicitly, we recover the solution 
\exref{eq:Fkm_final}. It is a simple matter to ascertain that \exref{eq:ssim_sln} 
has the asymptotics \exref{eq:Cm} and \exref{eq:Ck} when the similarity 
variable $k^3/9\gamma\tau$ is small or large, respectively. Indeed, 
changing the integration variable $z-k^3/9\gamma\tau \to -z$ in \exref{eq:ssim_sln}, 
we get 
\beq
F_k(\tau) = \frac{1}{\Gamma\!\lt(\frac{1}{3}\rt)}
\int_{0}^\infty \rmd z\,\frac{e^{-z}}{\bigl|\frac{1}{9\gamma}k^3 - \tau z\bigr|^{2/3}}
\to\lt\{
\begin{array}{ll}
\displaystyle \frac{1}{\tau^{2/3}},& 
\displaystyle \frac{|k|^3}{9\gamma\tau} \ll 1,\\\\
\displaystyle \frac{3^{4/3}}{\Gamma\!\lt(\frac{1}{3}\rt)}\frac{1}{k^2}, & 
\displaystyle \frac{|k|^3}{9\gamma\tau} \gg 1.
\end{array}
\rt.
\label{eq:ssim_limits}
\eeq
This solution manifestly has vanishing Hermite flux ($F_k=F_{-k}$) 
at $\tau\ll|k^3|/9\gamma$ (as well as in the opposite limit, but 
this is just a trivial requirement of continuity of $F_k$ at $k=0$). 

Obviously, the above derivation is much more elementary and may, to some readers, 
be more convincing than that offered in \secref{sec:solution}. 
Methodologically, however, it was not necessarily obvious either that 
the self-similar solution would prove to be the only one that 
would have matching ($k$ by $k$) phase-mixing and anti-phase-mixing fluxes 
at $\tau\to 0$ or that this solution would be the only physically acceptable one. 
As we already indicated above, the choice of $\al=2/3$ was merely a convenient 
conjecture. In order to ascertain that it is the only possible 
choice, we would have to solve \exref{eq:Phi} for arbitrary $\lambda$ 
(which can be done in special functions) and then demand that the solutions 
are positive (i.e., represent physically realisable spectra) and satisfy \exref{eq:zeroflux_tot}. 
This can be done, but the argument for $\al=2/3$ given in \apref{ap:inevitability}
is perhaps more mathematically compelling and at any rate no less cumbersome. 

Admittedly, the formal shortcomings of the self-similar route to the answer 
are probably overcome by its appealing simplicity. 

\section{Inevitability of the zero-flux solution \exref{eq:Y}}
\label{ap:inevitability}

Here we exercise due diligence by showing that the solution found in \secref{sec:solution}, 
which we found using an initial guess \exref{eq:Y}, is indeed 
the only physically and mathematically sensible one. 

Let us no longer make an explicit demand for a zero-flux solution, 
which led us from \exref{eq:flux_cont_sln} to \exref{eq:flux_cont_zeroflux}, 
the solution of which was \exref{eq:Y}. Instead of \exref{eq:Y}, let us 
posit some general power law for the Hermite spectrum at $\xx\to0$:  
\beq
Y(\tau) = \frac{1}{\tau^\al}.
\label{eq:Y_gen}
\eeq
With this ansatz, the left-hand side of \exref{eq:flux_cont_sln} becomes 
\begin{align}
\label{eq:lhs_gen}
\text{l.h.s. of \exref{eq:flux_cont_sln}} &= \tau^{-\al - \frac{1}{3}}
\Gamma\!\lt(\frac{2}{3}\rt) \lt[\frac{\Gamma\!\lt(\al+\frac{1}{3}\rt)}{\Gamma(\al)} 
- \frac{\lt(\al-\frac{2}{3}\rt)\Gamma(1-\al)}{\Gamma\!\lt(\frac{5}{3}-\al\rt)}\rt]\\
&\quad + \underbox{\frac{\rmd}{\rmd\tau}\,\tau^{\frac{2}{3}-\al} 
B\!\lt(\frac{\tau}{\tmax};\al-\frac{2}{3},\frac{2}{3}\rt)}{3cm}{$\to0$ as $\tmax\to\infty$},
\nonumber
\end{align}
where $B$ is the incomplete beta function. While the second integral 
in \exref{eq:flux_cont_sln}, equalling the expression under the derivative 
in \exref{eq:lhs_gen}, is divergent when $\al<2/3$ and $\tmax\to\infty$, 
its $\tau$ derivative vanishes in this limit and so the values $\al<2/3$ 
are formally allowed. The case $\al=2/3$ has to be treated differently, 
but has already been considered in \secref{sec:zeroflux}. Finally, 
we must have $\al<1$ in order to keep the first integral in \exref{eq:flux_cont_sln}
convergent. 

To calculate the right-hand side of \exref{eq:flux_cont_sln}, 
we assume a general power law for the phase-mixing part of the Fourier spectrum at $\tau\to0$, 
\beq
F_0^+(\xx) = \frac{A}{\xx^\be} = \lt(\frac{3}{2}\rt)^\be\!\frac{A}{k^{3\be/2}},  
\label{eq:Fp0_gen}
\eeq
where $A$ is a constant, and find 
\beq
\text{r.h.s. of \exref{eq:flux_cont_sln}} = 
\tau^{-\frac{\be}{2} - \frac{1}{3}}
2^{-\be}\Gamma\!\lt(1-\frac{\be}{2}\rt)A.
\eeq
In order for \exref{eq:flux_cont_sln} to be satisfied, we must have 
\beq
\al = \frac{\be}{2}
\label{eq:scalings}
\eeq
and 
\begin{align}
\nonumber
A &= 2^{2\al}\Gamma\!\lt(\frac{2}{3}\rt) 
\lt[\frac{\Gamma\!\lt(\al+\frac{1}{3}\rt)}{\Gamma(\al)\Gamma(1-\al)} 
- \frac{\al-\frac{2}{3}}{\Gamma\!\lt(\frac{5}{3}-\al\rt)}\rt]\\
&=  2^{2\al}\Gamma\!\lt(\frac{2}{3}\rt) 
\frac{\al-\frac{2}{3}}{\Gamma\!\lt(\frac{5}{3}-\al\rt)}
\lt\{\frac{\sin\pi\al}{\sin\!\lt[\pi\lt(\al - \frac{2}{3}\rt)\rt]}-1\rt\}.
\label{eq:A}
\end{align}
The spectrum \exref{eq:Fp0_gen} must be positive: within our allowed 
interval $\al<1$, we can have $A>0$ only if $\al<5/6$. 

Finally, via \exref{eq:Fm0}, the ansatz \exref{eq:Y_gen} implies that 
the anti-phase-mixing Fourier spectrum at $\tau\to 0$ is 
\beq
F_0^-(\xx) = 2^{2\al}\frac{\Gamma\!\lt(\al+\frac{1}{3}\rt)}{\Gamma\!\lt(\frac{1}{3}\rt)}
\frac{1}{\xx^{2\al}}. 
\label{eq:Fm0_gen}
\eeq
(the integral is done by changing the integration variable to $z=\xx^2/4\sigma$). 
In view of \exref{eq:scalings}, this is the same scaling as for the 
phase-mixing spectrum \exref{eq:Fp0_gen}! 
But this means that the condition \exref{eq:zeroflux_tot} can only 
be satisfied if the coefficients in \exref{eq:Fm0_gen} and \exref{eq:Fp0_gen}
match. With $A$ given by \exref{eq:A}, this leads us, therefore, to demand  
\beq
A = 2^{2\al}\frac{\Gamma\!\lt(\al+\frac{1}{3}\rt)}{\Gamma\!\lt(\frac{1}{3}\rt)}
\quad
\Leftrightarrow
\quad
\cos\!\lt[\pi\!\lt(\al-\frac{1}{3}\rt)\rt] = \frac{1}{2}.
\label{eq:Amatch}
\eeq 
This is satisfied for $\al=2/3$, which hands us back our zero-flux 
solution \exref{eq:Y} and everything that it implies. 

The reason that this has proved to be the only physically acceptable possibility 
is that for any $0<\al<2/3$, we would have $F_0^+ > F_0^-$ and their 
wave-number scaling would be shallower than $1/k^2$, causing 
the integral in \exref{eq:zeroflux_tot} to blow up. For $2/3<\al<5/6$, 
the integral is finite but negative, meaning that there is no steady state 
and there is an unphysical net inflow of energy from high $m$.
For even larger values of $\al$, either there is no positive 
spectrum at all or the energy flux through $k=0$ blows up. 
Finally, $\al = 0$ is not allowed because, even though \exref{eq:Amatch} 
and so \exref{eq:zeroflux_tot} are satisfied, the right-hand side 
of \exref{eq:budget_stst} no longer vanishes at $\nu\to+0$ 
and so the steady-state energy budget breaks down, i.e., no such 
steady-state solution can exist---this is a demonstration 
that the linear Hermite solution does not work in a nonlinear steady state.   

\end{appendix}

\bibliography{../JPP/bib_JPP}{}
\bibliographystyle{jpp}

\end{document}